\documentclass[journal,twocolumn,final]{IEEEtran}
\usepackage[pdftex]{graphicx}
\usepackage[cmex10]{amsmath}
\usepackage{graphicx}
\usepackage{epstopdf}
\usepackage{xcolor}
\usepackage{cite} 
\usepackage{flushend}
\usepackage{amssymb}
\usepackage{amsfonts}
\usepackage{multirow}
\usepackage[hidelinks]{hyperref}
\usepackage{algorithm}
\usepackage{algorithmic}
\usepackage{soul}
\usepackage{bm}

\begin{document}
\title{Ensemble Nonlinear Model Predictive Control for Residential Solar--Battery Energy Management}
\author{\normalsize Yang~Li,~\IEEEmembership{Member,~IEEE,}
     D.~Mahinda~Vilathgamuwa,~\IEEEmembership{Fellow,~IEEE,}
     Daniel~E.~Quevedo,~\IEEEmembership{Fellow,~IEEE,}
     Chih~Feng~Lee,
       and Changfu~Zou,~\IEEEmembership{Senior Member,~IEEE}    
\thanks{Yang Li and Changfu Zou are with the Department of Electrical Engineering, Chalmers University of Technology, Gothenburg 41296, Sweden (e-mail: yangli@ieee.org; changfu.zou@chalmers.se).}
\thanks{D. Mahinda Vilathgamuwa and Daniel E. Quevedo are with the School of Electrical Engineering and Robotics, Queensland University of Technology, Brisbane, QLD 4001, Australia (e-mail: mahinda.vilathgamuwa@qut.edu.au; daniel.quevedo@qut.edu.au).}
\thanks{Chih Feng Lee is with Polestar, Gothenburg 41878, Sweden (e-mail: chih.feng.lee@polestar.com).}
}

\maketitle

\begin{abstract}
In a dynamic distribution market environment, residential prosumers with solar power generation and battery energy storage devices can flexibly interact with the power grid via power exchange. Providing a schedule of this bidirectional power dispatch can facilitate the operational planning for the grid operator and bring additional benefits to the prosumers with some economic incentives. However, the major obstacle to achieving this win-win situation is the difficulty in 1)  predicting the nonlinear behaviors of battery degradation under unknown operating conditions and 2) addressing the highly uncertain generation/load patterns, in a computationally viable way. This paper thus establishes a robust short-term dispatch framework for residential prosumers equipped with rooftop solar photovoltaic panels and household batteries. The objective is to achieve the minimum-cost operation under the dynamic distribution energy market environment with stipulated dispatch rules. A general nonlinear optimization problem is formulated, taking into consideration the operating costs due to electricity trading, battery degradation, and various operating constraints. The optimization problem is solved in real-time using a proposed ensemble nonlinear model predictive control-based economic dispatch strategy, where the uncertainty in the forecast has been addressed adequately albeit with limited local data. The effectiveness of the proposed algorithm has been validated using real-world prosumer datasets.
\end{abstract}


\bstctlcite{BSTcontrol}

\section{Introduction} \label{sec:sec_1}

\IEEEPARstart{S}{olar} photovoltaic (PV) systems have become increasingly prevalent for residential users worldwide in recent years \cite{irena2019a}.
A case in point is that one-fourth of Australian households have installed rooftop PV panels by the end of 2020 \cite{csiro}.
As the penetration of such distributed PV generation increases, the intermittent and stochastic nature of solar irradiance has started to negatively impact the operation of the distribution network in various ways.
Although utilizing these free-of-charge energy can bring some expected economic benefits, these benefits are often offset by power quality and stability issues.
Integration of energy storage systems for prosumers has become an economically favorable solution to mitigate these issues, particularly with the ever-declining prices of lithium-ion (Li-ion) batteries. This is especially true when the distribution system operator (DSO) permits more flexible bi-directional electricity transactions.
In this context, a well-designed power exchange strategy in the home energy management system (HEMS) can effectively reduce the overall operating costs \cite{kong2021tsg,ito2018tcst}.

A plethora of research studies have investigated the design of control strategies for HEMS \cite{gomes2022energies}.
Initial approaches were  rule-based techniques to achieve computational simplicity, e.g., those based on self-consumption maximization or  time-of-use arbitrage \cite{azuatalam2018rser}.
While these algorithms are easy to implement, they are difficult to generalize and are typically suboptimal.
Optimization-based HEMS strategies have become the focus of later investigations, with the primary control objective of minimizing overall costs while adhering to various operational constraints.
In \cite{bozchalui2012a}, different types of household appliance models were established for optimizing energy consumption, but the authors did not consider bi-directional power flow, which allows the consumer to sell electricity to the utility grid.
In \cite{erdinc2015a}, a bi-directional plug-in electric vehicle charging/discharging model was proposed where the degradation effect of the battery was ignored.
In \cite{wu2016a}, the authors proposed a nonlinear predictive strategy for a residential building equipped with PV panels and batteries, with the aim of  minimizing the operating cost  of the system under a bidirectional grid power setting. The study highlighted the importance of considering electricity price, battery degradation, and emission level.

When the goal of HEMS is solely to maximize the prosumer's own benefit without considering grid conditions, as in previous studies \cite{azuatalam2018rser,bozchalui2012a,erdinc2015a,wu2016a}, the optimization results may lead to conflicting and detrimental effects on the distribution network if multiple users are using the same optimization algorithm.
To address this problem, recent literature suggests utilizing multi-objective optimization and coordinated control schemes \cite{saberi2021tsg}.
For example, in \cite{nizami2018}, a multi-objective optimization-based framework was proposed to determine the charge-discharge schedule of residential batteries to reduce electricity bills.
Additionally, \cite{mahmud2016} investigated the capability of HEMS to improve peak shaving using a similar configuration of PV, battery storage, and electric vehicle.
To match the residential load demand with small-scale renewable energy sources, an intelligent multi-objective home energy management scheme was proposed in \cite{lokeshgupta2019}.
This scheme can minimize the prosumer's energy bill and the system peak demand simultaneously.

The above-mentioned multi-objective optimization-based control requires the coordination of a large number of users with effective while computationally expensive communication and a large amount of data, subject to the prosumer's willingness to share the local information.
For privacy reasons, these requirements can hinder the practical implementation of the algorithms.
A  technically feasible and cost-effective solution is to emulate the conventional day-ahead and intraday market mechanisms used for large centralized generation units, wherein every residential unit submits a power generation schedule to the DSO \cite{liyang2020apen}.
In this scenario, the future power exchange schedule at the point of connection (PoC) of a prosumer shall be provided in advance. This can facilitate the operational planning of the distribution network within a specific period with stipulated parameters.
However, designing a short-term dispatch strategy for HEMS can be challenging due to the difficulty in predicting the multi-timescale and multi-physics nonlinear dynamics of Li-ion batteries. These dynamics are affected by highly uncertain degradation, environmental conditions, energy market conditions, and load patterns over a long prediction horizon.

The \emph{battery model} plays a critical role in selecting the optimization method of HEMS.
Conventionally, algorithms such as linear programming or mixed-integer linear programming were used with assumed linearized battery characteristics and simple or even no degradation dynamics \cite{bozchalui2012a,erdinc2015a,hou2019}.
Unfortunately, linear optimization results can deviate significantly from the true optimal solution when applied to a real battery with highly nonlinear and complex degradation mechanisms, especially over long prediction horizons.
Global solutions can be obtained through the use of nonlinear optimization methods such as deterministic dynamic programming \cite{sun2016a}, particle swarm optimization, binary particle swarm optimization \cite{dinh2020}, and fuzzy logic \cite{DERROUAZIN2017238}.
However, efficiently implementing these algorithms for online operation can be challenging due to their high computational overheads and sensitivity to parameter changes.
In contrast, model predictive control (MPC) is a practical and effective method widely adopted  for HEMS \cite{sun2016a,wu2016a,ito2018tcst}, battery charging \cite{pozzi2020tcst}, and microgrid \cite{pippia2020tcst}, provided that an accurate system model is available.
Unfortunately, due to a lack of domain knowledge and the need for low computational complexity, the battery models used in the studies mentioned above are often empirically established by fitting laboratory data based on parameters such as state-of-charge, depth-of-discharge, temperature, voltage, and cycle number, among others \cite{azuatalam2018rser}.

However, it is well-known that empirical battery models cannot be readily extrapolated to an operating condition different from the experimental design \cite{reniers2021jps}. Given that optimal control strategies are determined online rather than in advance, the use of empirical battery models without knowledge of operational insights is deemed unreliable, especially for predicting battery degradation behaviors affected by various internal factors not reflected in such models.
The concept of incorporating a physics-based battery model into HEMS was investigated\cite{liyang2018iecon}. However, this model was developed for isothermal conditions by assuming that the battery temperature was ideally maintained at a reference value. Since temperature significantly affects battery performance, this assumption prevents accurate estimations of overall battery performance under dynamic environmental conditions.

On the other hand, \emph{forecast uncertainty} in the HEMS is usually addressed by assuming an error bound, and the optimal control strategy is typically determined using only the mean (or nominal) nonlinear MPC (mNMPC) \mbox{\cite{liyang2018iecon,CAI2019478}}.
Because economic operations usually drive the system to its constraints, forecast uncertainty becomes very important. However, the robustness of these algorithms has not been thoroughly addressed, and the potential impact of using mNMPC on battery degradation has not been fully explored.
Stochastic dynamic programming \mbox{\cite{wu2016a}} was used to account for the randomness in user demand, but its implementation necessitates a vast amount of historical data and substantial parallel computing resources to identify the load paradigm. See, e.g., \cite{mesbah2016mcs} for a systematic review of stochastic MPC (SMPC). SMPC algorithms such as chance-constrained MPC \cite{garifi2018} and non-conservative robust MPC (RMPC) such as multistage MPC \cite{lucia2014jpc} can also be incorporated into this framework. Nevertheless, the current nonlinear battery model presents challenges for online calculations due to difficulty in handling probabilistic uncertainty with tractability \cite{moura2013tcst}. Initial efforts have been focused on addressing the challenges associated with SMPC and RMPC by introducing ensemble MPC (EnMPC) \cite{boiroux2016ecc} and ensemble nonlinear MPC (EnNMPC) \cite{garcia2019acc} for blood glucose control. These methods strive to achieve a balance between computational complexity and uncertainty levels. Nonetheless, the method needs to generate disturbance samples using physics-informed models and storing a significant amount of data, which may not be economically or technically viable for HEMS applications.
Therefore, a robust and easy-to-implement framework that can efficiently provide a power exchange schedule for the economic dispatch of prosumer energy for both day-ahead and intraday operations would be highly beneficial.

To overcome the challenges identified above, a novel economic dispatch control framework for HEMS is proposed that enables residential prosumers to submit their desired buying or selling schedules to the DSO. There are two main contributions compared to the above-mentioned works. First, the dynamics of Li-ion batteries are incorporated into the proposed framework using a third-order nonlinear physics-informed model, which considers temperature effects and degradation. This model can be used for the cost-benefit evaluation and the corresponding control. The physical descriptions of battery dynamics ensure high model accuracy under arbitrary battery power flow. The accuracy of the model is verified by comparing it with a high-fidelity pseudo-two-dimensional electrochemical model. The result demonstrates that the proposed model captures electrical, thermal, and aging behaviors with high precision, while requiring very low model complexity. Second, motivated by the EnNMPC, a novel ensemble-based economic dispatch strategy is proposed to solve a stochastic optimization problem. Furthermore, the statistical characteristics of the past forecast error behaviors are learned using only a small amount of  historical data, regardless of disturbance models. With the aim that the scheme should be generally applicable but not subject to specific resources and techniques for forecasting, an ensemble of forecast errors are generated based on a small amount of recent data, thereby requiring minimal computation and data storage.

The remainder of this work is organized as follows. Section~\ref{sec:sec_2} presents the system configuration of the HEMS and the market rule under investigation, based on which the general problem statement for the optimal HEMS dispatch is established. Section~\ref{sec:sec_3} develops the model for the HEMS, followed by a detailed description of the control objective, the constraints, and the algorithm of the proposed EnNMPC-based energy management strategy in Section~\ref{sec:sec_4}. Results are presented and discussed in Section~\ref{sec:sec_5} to show the effectiveness of the proposed approach. Finally, the concluding remarks are given in Section~\ref{sec:sec_6}.

\section{System Description} \label{sec:sec_2}

\subsection{System Configuration}  \label{sec:sec_2a}

\begin{figure}[!tb]
  \centering
  \includegraphics[width=0.48\textwidth]{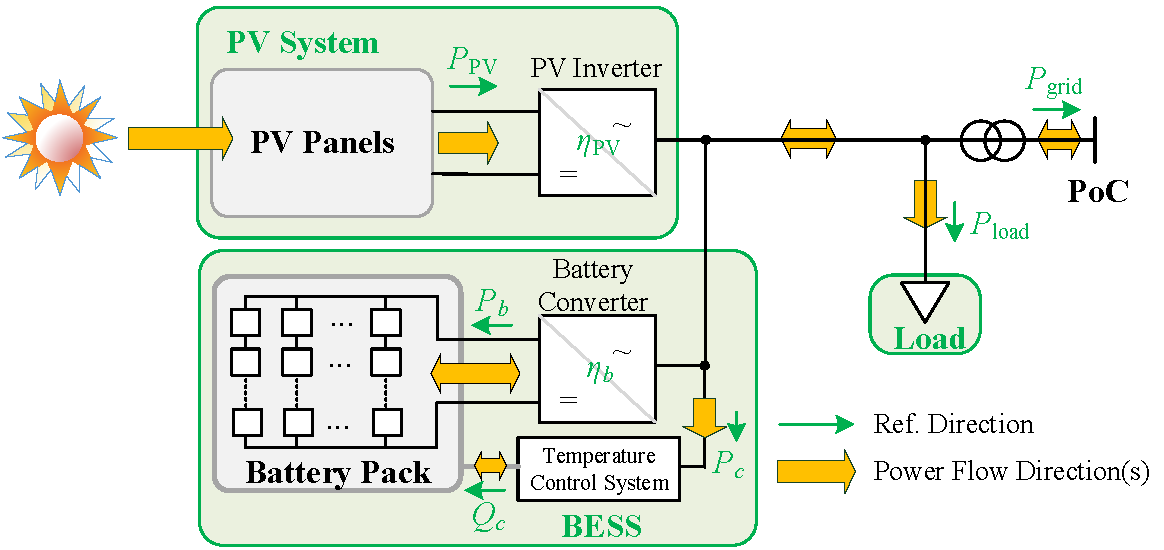}
  \caption{Typical residential prosumer configuration with PV and batteries.}
  \label{fig:fig_1}
\end{figure}

A typical prosumer configuration with rooftop PV panels and Li-ion batteries is shown in Fig.~\ref{fig:fig_1}. The PV panels and the Li-ion battery pack are connected to the utility grid at the PoC via the PV inverter and the battery system converter, respectively.
The PV is configured with a maximum power point tracking scheme to ensure the full utilization of solar energy.
$P_{\text{PV}}$ and $P_{b}$ represent the output power of the PV panels and the charging power of the battery pack, respectively, and $\eta_{\text{PV}}$ and $\eta_{{b}}$ represent the efficiencies of the corresponding power converters.
This work specifically considers the battery temperature, which must be maintained within an appropriate operating range. The electrical power consumption of the temperature control system is denoted by $P_c$.
The net difference between the combined PV/battery systems and the load $P_{\text{load}}$ is compensated by exchanging power $P_{\text{grid}}$ with the utility grid at the PoC via a distribution transformer.
It is worth noting that other control degrees of freedom can be readily integrated into this configuration framework, such as PV curtailment, deferrable loads, and bi-directional chargers of electric vehicles. For the sake of brevity, these components are not specifically modeled in this work; however, their impacts are assumed to be accounted for in the highly random load.  Furthermore, Fig.~\ref{fig:fig_1} can be generalized to represent  the larger context of a residential building or community where the users share a central battery bank, while the PV and load represent their aggregated values.

\subsection{Dispatch Rules and Notations} \label{sec:sec_2b}

\begin{figure}[!t]
\centering
  \includegraphics[width=0.48\textwidth]{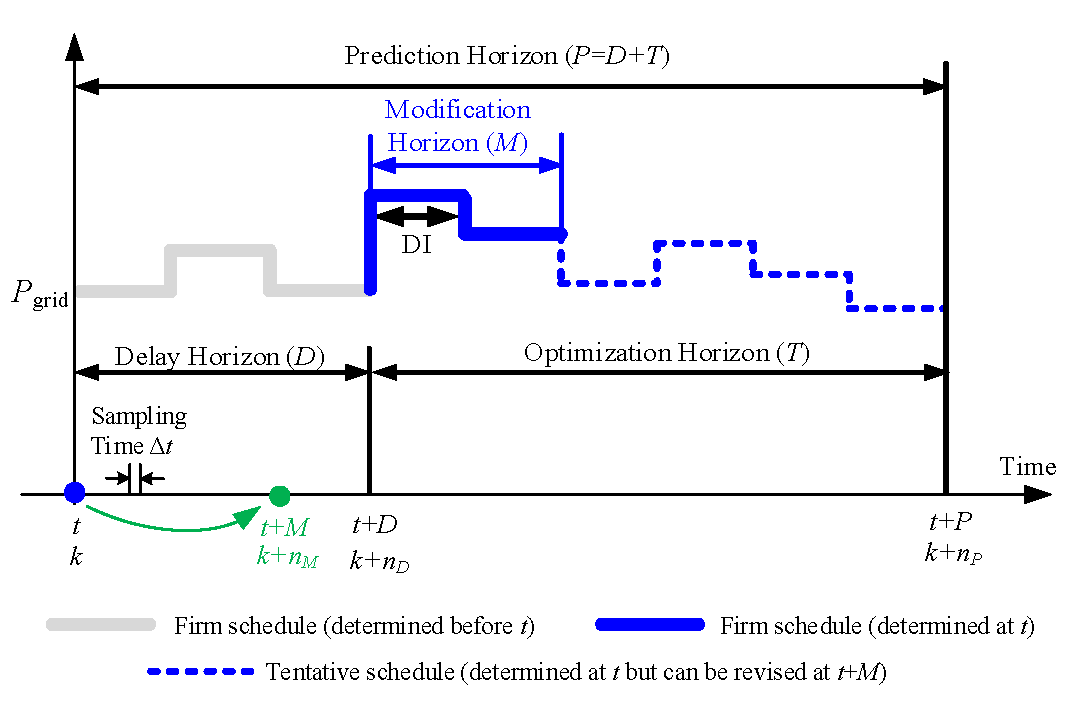}
  \caption{Example dispatch rules for prosumers under investigation, where $D = 3 (\text{DI})$,  $M = 2 (\text{DI})$,  $T = 6 (\text{DI})$.}
  \label{fig:fig_2}
\end{figure}

In this work, we will present our home energy dispatch algorithm based on the dispatch rules described as follows. These rules are similar to that applicable to large-scale renewable generating units for achieving short-term dispatchability \cite{liyang2020apen}. According to the rules illustrated in Fig.~\ref{fig:fig_2}, the prosumer is to submit the $T$-hour dispatch schedule $D$ hours ahead of time. The schedule consists of a sequence of \emph{dispatch intervals} (DIs) of $P_{\text{grid}}$. The prosumer is allowed to modify and resubmit the schedule every $M$ hours except for the immediate $D$ hours. As shown in Fig.~\ref{fig:fig_2}, at time $t$, the schedule of $P_{\text{grid}}$ to be submitted is from $t+D$ to $t+D+T = t+P$, while the schedule of $P_{\text{grid}}$ from $t$ to $t+D$ has been determined previously as a firm commitment and is not subject to adjustment. In this work, $D$ is referred to as the \emph{delay horizon}, $T$ is the \emph{optimization horizon}, $M$ is called the \emph{modification horizon}, and $P = D+T$ represents the \emph{prediction horizon}.

In the rest of the work, we will investigate the short-term dispatch problem in the discrete-time domain, where the sample time $\Delta t$ is considered smaller than a DI. A small $\Delta t$ can facilitate more accurate prediction of the violation of operating constraints and the nonlinear degradation behaviors, while $\Delta t$ should large enough to ensure affordable computation.
A subscript $k \in \{0,1,\cdots \}$ is attached to a variable as the index of the present sampling instant $t = k \Delta t$. The following notations are adopted:
\begin{align}
n_D &:= D/\Delta t, \quad n_T := T/\Delta t, \quad n_M := M/\Delta t, \nonumber \\
n_P &:= P/\Delta t, \quad n_\text{DI}:= (\text{DI})/\Delta t,\nonumber\\
\mathcal{S}_D &:= \{k, \cdots, k + n_D - 1\},\nonumber \\
\mathcal{S}_T &:= \{k + n_D, \cdots, k + n_P - 1\},\nonumber \\
\mathcal{S}_M &:= \{k + n_M, \cdots, k + n_M - 1\},\nonumber \\
\mathcal{S}_P &:= \mathcal{S}_D \cup \mathcal{S}_T = \{k, \cdots, k + n_P - 1\}. \nonumber
\end{align}

Furthermore, throughout this work, time, electric charge or charge capacity, and power quantities are expressed in units of hour (h), ampere-hour (Ah), and kilowatt (kW), while voltage, current, and resistance are expressed in units of volt (V), ampere (A), and ohm ($\Omega$), respectively.

\subsection{General Problem Statement} \label{sec:sec_2c}

According to the system configuration and the dispatch rules described in the previous two subsections, the objective of the short-term dispatch for the HEMS shall be: At time instants $t \in \{0, M, 2M \cdots\}$, determine the power exchange profile $P_{\text{grid}}$ for a future period $(t+D, t+P]$ which gives the least cost of operation. The power exchange profile should ensure that system variables such as battery current, voltage, state-of-charge (SoC), and local power flows, are well-maintained within respective allowable operating ranges. Following industry practices, we shall further assume that the power exchange schedule $P_{\text{grid}}$ within a period $(t, t+D]$ as well as the forecasts of the PV power, local load, and electricity price are available for the prosumer. We formulate this general problem description into a constrained finite-horizon optimization problem by the following. At discrete time instant $k$ (which should be the first instant of a DI), given 1) the estimated initial state $\hat{x}_{k}$, where $\hat{ }$ denotes the estimated value, and 2) the pre-determined dispatch schedule

${{{\mathbf{P}}_{\text{grid},k}^{\prime}}}(\cdot) := [P_{\text{grid},k}, P_{\text{grid},k  + n_\text{DI}},  \cdots, P_{\text{grid},k  + n_D - n_\text{DI}}] \in \mathbb{R}^\frac{D}{{\text{DI}}},$

and 3) the disturbance sequence
\begin{equation}{\mathbf{d}}_{k:k + n_P - 1}(\cdot) := [d_k, d_{k+1}, \cdots  d_{k + n_P - 1}] \in \mathbb{R}^{n_d \times n_P}, \nonumber
\end{equation}
seek the short-term dispatch schedule
\begin{align}{{{\mathbf{P}}_{\text{grid},k}}}(\cdot) := & [P_{\text{grid},k + n_D}, P_{\text{grid},k + n_D + n_\text{DI}},  \cdots, \nonumber \\
 &P_{\text{grid},k + n_D + (\frac{T}{{{\text{DI}}}}-1) n_\text{DI}}] \in \mathbb{R}^\frac{T}{{\text{DI}}}, \nonumber
\end{align}
to achieve
\begin{subequations} \label{eq:eq_1}
    \begin{align} \label{eq:eq_1a}
    \min_{{{\mathbf{P}}_{{\text{grid}},k}}(\cdot), v_k} \quad & \sum_{i \in \mathcal{S}_T}{\text{Cost}^{\text{op}}_i} \\
   \textrm{s.t.} \quad & x_k = \hat{x}_{k}  \label{eq:eq_1b}\\
   & x_{i + 1}=f(x_{i}, u_{i} ,{d}_{i}) \quad \forall i \in \mathcal{S}_P \label{eq:eq_1c} \\
   & y_{i} =h(x_{i},u_{i}, {d}_{i})  \quad \forall i \in \mathcal{S}_P \label{eq:eq_1d} \\
   & x_{i + 1} \in \mathcal{X} \subset {{\mathbb{R}}^{n_x}}  \quad \forall i \in \mathcal{S}_P  \label{eq:eq_1e} \\
   & y_{i} \in \mathcal{Y} \subset {{\mathbb{R}}^{n_y}}  \quad \forall i \in \mathcal{S}_P \label{eq:eq_1f} \\
   & {{u}_{i}} =  \mathcal{U} \subset \mathbb{R}^{n_u}  \quad \forall i \in \mathcal{S}_P \label{eq:eq_1g} \\
   & P_{\text{grid},i} = \left.P_{\text{grid},l}\right|_{l =\lfloor {i}/{n_\text{DI}} \rfloor n_\text{DI}}  \quad \forall i \in \mathcal{S}_P \label{eq:eq_1h}
    \end{align}
\end{subequations}

In the above, $\text{Cost}^{\text{op}}$ is the operating cost per unit time (i.e., the running cost), $f: {{\mathbb{R}}^{n_x}} \times {{\mathbb{R}}}^{n_u} \times {{\mathbb{R}}^{n_d}} \rightarrow {{\mathbb{R}}^{n_x}}$  and  $h: {{\mathbb{R}}^{n_x}} \times {{\mathbb{R}}}^{n_u} \times {{\mathbb{R}}^{n_d}} \rightarrow {{\mathbb{R}}^{n_y}}$ are two nonlinear maps for the process and the output of the prediction model. $\mathcal{X}$, $\mathcal{Y}$, and $\mathcal{U}$ define the operating constraints for the state $x$, output $y$, and input $u$, respectively. The control input ${{u}} := [P_{\text{grid}}, v]^{\top}$ in \eqref{eq:eq_1g} consists of the dispatch power and other control input $v$. In \eqref{eq:eq_1h}, a floor function $\lfloor \cdot \rfloor$ is used to indicate that only the first controlled input in each DI is considered the decision variable in ${{{\mathbf{P}}_{{\text{grid}},k}}}(\cdot)$. Consequently, the computational complexity of the problem can be significantly reduced compared to the case when the entire future input sequence ${\mathbf{u}}_{k + n_D:k + n_P - 1}(\cdot) := [u_{k+ n_D}, u_{k+ n_D+1}, \cdots  u_{k + n_P - 1}] \in \mathbb{R}^{n_u \times n_T}$ is selected as the decision variables: The number of decision variables is reduced by a factor of $(n_u n_\text{DI})/((n_u-1) n_\text{DI} + 1)$. Since the future disturbances ${\mathbf{d}}_{k:k + n_P - 1}(\cdot)$ are not known exactly, they have to be forecasted for solving the optimization problem \eqref{eq:eq_1}.
The prediction model, operating constraints, operating costs, and disturbances will be discussed in the next two sections.

\section{Prediction Model} \label{sec:sec_3}

\subsection{Physics-Based ECM for Li-Ion Batteries} \label{sec:sec_3a}

Two isothermal degradation-conscious physics-based equivalent circuit models (PB-ECMs) were developed based on the isothermal electrochemical single particle model of the Li-ion battery for operational planning in our previous work, including a 4th-order and a 2nd-order models \cite{liyang2019apen}. The 4th-order PB-ECM requires a small sampling time for obtaining the solution as it possesses fast dynamics with a time constant in the order of several minutes. This is acceptable for designing a nominal MPC-based dispatch strategy \cite{liyang2019igbsg}, but can lead to a robust MPC algorithm intractable for online operation since many more scenarios need to be evaluated within each optimization period. Hence, we adopt the 2nd-order PB-ECM and consider the temperature influence:
\begin{subequations} \label{eq:eq_2}
    \begin{align}
   \dot{Q}_{1}^{\pm}(t)&= I_{1}^{\pm}(t) \label{eq:eq_2a}\\
    I_{\text{1}}^{+}(t)&={{I}_{\text{bat}}(t)} \label{eq:eq_2b} \\
    I_{\text{1}}^{-}(t)&=-{{I}_{\text{bat}}(t)}-I_{\text{sr}} \label{eq:eq_2c} \\
{{V}_{\text{bat}}(t)}& = {{V}_{1}^{+}}  + R_{\Sigma}^+ I_{1}^+(t) -{{V}_{1}^{-}} -  R_{\Sigma}^- I_{1}^-(t) \nonumber \\& + [R_{f}(t) + R_e  + R_{\text{col}}]{{I}_{\text{bat}}(t)} \label{eq:eq_2d}
    \end{align}
where $I_{\text{bat}}$ is the  cell current and $V_{\text{bat}}$ is the cell terminal voltage. $Q_1$, $V_1$, and $I_1$ are the electric charge, the open-circuit potential, and the main-reaction current of an electrode, respectively. The superscript $\pm \in \{+, -\}$ where ``$+$'' and ``$-$'' represent the positive and the negative electrodes, respectively.  Furthermore, $I_{\text{sr}}$ is the side reaction current, $R_\Sigma$ is the sum of charge-transfer resistance and solid-phase diffusion resistance, $R_f$ is the solid-electrolyte interphase (SEI) resistance, $R_e$ is the electrolyte resistance, and $R_\text{col}$ is the resistance of the current collectors. Note that $V_{\text{bat}}$, $R_\Sigma$, $I_{\text{sr}}$, and $R_e$ are expressed as nonlinear functions (see details in Appendix)
    \begin{align}
V_1^{\pm} &= {{f}^{\pm}_{V_1}}(Q_1^{\pm}, {T}_{\text{bat}}) \label{eq:eq_2e} \\
R_{\Sigma}^{\pm} &= {{f}^{\pm}_{R_\Sigma}}(Q_1^{\pm}, {T}_{\text{bat}}) \label{eq:eq_2f} \\
I_{\text{sr}} &= f_{I_{sr}}(Q_1^{-}, {T}_{\text{bat}}, I_{\text{bat}}) < 0 \label{eq:eq_2g}
\\
R_{{e}} &= f_{R_{e}}({T}_{\text{bat}}).
 \label{eq:eq_2h}    \end{align}
\end{subequations}

The relationship between battery pack power $P_b$, cell current, cell voltage, and cell number $N_{\text{cell}}$ is
\begin{equation} \label{eq:eq_3}
{{P}_{b}(t)}={{N}_{\text{cell}}}{{V}_{\text{bat}}(t)}{{I}_{\text{bat}}(t)}\times {{10}^{-3}} \quad [\text{kW}].
\end{equation}

It should be noted that \eqref{eq:eq_2} and \eqref{eq:eq_3} is a system of differential-algebraic equations if the battery power $P_b$ is considered as an input variable. To avoid using computationally inefficient iterative methods for solving such a system, we plug \eqref{eq:eq_2b}--\eqref{eq:eq_2d} into \eqref{eq:eq_3}, and considering $|I_{\text{sr}}| \ll |I_{\text{bat}}|$ and $I_{\text{sr}} \approx 0$, the battery current can be solved as
\begin{equation} \label{eq:eq_4}
{{I}_{\text{bat}}}(t) = (-V_{\text{eq}} + \sqrt{V_{\text{eq}}^2 + 4 R_{\text{eq}} [P_b(t)/N_\text{cell} \times 10^3}])/(2 R_{\text{eq}})
\end{equation}
where $V_{\text{eq}} =  {{V}_{1}^{+}}-{{V}_{1}^{-}}$ and $R_{\text{eq}} = {{R}_{1}^{+}}+{{R}_{1}^{-}} +  {R}_{f}+{R}_{e}+{R}_\text{col}$ are the equivalent voltage and the equivalent resistance of the battery cell, respectively.

\subsection{Battery Degradation Model}

The irreversible side reactions occurring in the negative electrode are considered for the degradation model.
The side reactions lead to two degradation phenomena, i.e., capacity fade and increase of the internal resistance.
First, from \eqref{eq:eq_2a}--\eqref{eq:eq_2c} and \eqref{eq:eq_2g}, it can be seen that since the side reaction current ${I}_{\text{sr}}$ is negative, the sum $Q_{1}^{+}(t) + Q_{1}^{-}(t) = Q_0 -\int_0^t {I}_{\text{sr}}(\tau) d \tau$ will keep increasing over time, where ${{Q}_{0}} = Q_{1}^{+}(0) + Q_{1}^{-}(0)$ is the capacity for a fresh cell. This amount of ``unbalanced'' capacity change will cause a part of capacity to be unusable and we denote it as the capacity loss ${Q}_{\text{loss}}$, i.e,
    \begin{equation} \label{eq:eq_5}
    {{Q}_{\text{loss}}}(t) = -\int_0^t {I}_{\text{sr}}(\tau) d \tau = Q_{1}^{+}(t) + Q_{1}^{-}(t) - {{Q}_{0}} \ge 0
    \end{equation}

Note that in \eqref{eq:eq_5} the equality condition  only holds for a fresh cell when both $Q_{1}^{+}$ and $ Q_{1}^{-}$ vary within the range $[0, {{Q}_{0}}]$. When the battery gets aged, the increase of the sum $Q_{1}^{+} + Q_{1}^{-}$ will lead to an increase in ${{Q}_{\text{loss}}}$, and both the ranges of $Q_{1}^{+}$ and $ Q_{1}^{-}$ are narrowed down from $[0, {{Q}_{0}}]$ to $[{{Q}_{\text{loss}}}, {{Q}_{0}}]$. We can thus define the state-of-health (SoH) of the battery according to the capacity loss:
    \begin{equation} \label{eq:eq_7}
    \text{SoH}(t) = \frac{{{{Q}_{0}}-{{Q}_{\text{loss}}(t)}}}{Q_{0}} = 1-\frac{{Q}_{\text{loss}}(t)}{{Q}_{0}}.
    \end{equation}

The SoC, used to indicate the level of stored energy, is defined according to the remaining electric charge in the electrode, i.e.,
\begin{equation} \label{eq:eq_8}
\text{SoC}(t)=\frac{{Q_{1}^{+}(t) - {{Q}_{\text{loss}}}}}{{{{Q}_{0}}-{{Q}_{\text{loss}}(t)}}}=\frac{{{{Q}_{0}}-Q_{1}^{-}(t)}}{{{{Q}_{0}}-{{Q}_{\text{loss}}(t)}}}.
\end{equation}

On the other hand, the SEI film growth in the negative electrode contributes to the gradual increase of the film resistance $R_{f}$ and thus the overall internal resistance. The change in $R_{f}$ is considered to be proportional to ${Q}_{\text{loss}}$, i.e.,
    \begin{equation} \label{eq:eq_6}
    R_{f}(t) = R_{f0} + K_f {{{Q}}_{\text{loss}}(t)}
    \end{equation}
where $R_{f0}$ is the SEI film resistance of a fresh cell and $K_f$ is a constant to describe the linear dependence between the increase of resistance and capacity loss, and their expressions are given in Appendix. The reader is referred to \cite{liyang2019apen} for a more detailed explanation of the capacity fade and the SEI resistance growth phenomena due to the side reactions of Li-ion cells.

\subsection{Battery Temperature Model} \label{sec:sec_3c}

In order to reflect the temperature effect on the battery pack and the corresponding control, we incorporate a lumped thermal model \cite{rosewater2020}:
\begin{equation}
C_T \dot{T}_{\text{bat}}(t) = (T_{\text{amb}}(t) - T_{\text{bat}}(t))/R_T + Q_\text{{gen}}(t) + Q_c(t)
\end{equation}
where $C_T$ and $R_T$ are the thermal capacitance and thermal resistance of the battery system, respectively, $T_{\text{amb}}$ is the ambient temperature, and $Q_c$ is the heating/cooling power used to regulate the battery temperature. We consider the electric power consumption as
\begin{equation}
P_c(t) = |Q_c(t)|/\eta_c
\end{equation}

\begin{subequations}
The heat generation $Q_\text{gen}(t)$ consists of two components namely the irreversible heat and the reversible heat:
\begin{align}
Q_\text{{gen}}(t) =Q_\text{{irrev}}(t) + Q_\text{{rev}}(t)
\end{align}

The irreversible heat is the result of the Joule effect. It can be calculated according to the resistive loss of the physics-based ECM:
\begin{align}
Q_\text{{irrev}}(t) =N_{\text{cell}} {I}_{\text{bat}}^2(t) R_{\text{eq}} \times 10^{-3}
\end{align}

On the other hand, the reversible heat is due to the entropy change:
\begin{align}
Q_\text{{rev}}(t) = N_{\text{cell}} {I}_{\text{bat}}(t) T_{\text{bat}}(t) \left(K_T^+ - K_T^-\right) \times 10^{-3}
\end{align}
where $K_T^+$ and $K_T^-$ are the entropic coefficients of the two electrodes, respectively. This reversible heat is important to accurately predict heat generation, especially in applications with low current rates \cite{YOO2019226715}, but is usually neglected in the existing literature for home energy management \cite{rosewater2020}. We will demonstrate its significance in Section V-A.
\end{subequations}

\subsection{Power Balance and Model Summary} \label{sec:sec_3d}
According to the system configuration shown in Fig.~\ref{fig:fig_1}, the power balance of the HEMS is governed by
\begin{equation} \label{eq:eq_12}
{{\eta }_{\text{PV}}}{{P}_{\text{PV}}(t)} - [{P}_{b}(t)/\eta _{b} + P_c(t)] ={{P}_{\text{grid}}(t)}+{{P}_{\text{load}}(t)}.
\end{equation}

The parametric values of the efficiencies of the converter depend on various factors such as the level and direction of power flow, voltage level, switching frequency, etc. However, they are simply considered constant for the given direction of power flow since the influence of the power electronic converter on system performance is not the focus of the present work.

The resulting continuous-time model \eqref{eq:eq_2} and \eqref{eq:eq_4}--\eqref{eq:eq_12} is next discretized using the forward Euler method and expressed in the state-space form \eqref{eq:eq_1c} and \eqref{eq:eq_1d}. Here, ${x}_i = [Q_{1,i}^+, Q_{1,i}^-, T_{{\text{bat}},i}]^{\top}$, ${y}_i = [\text{SoC}_i, \text{SoH}_i, V_{\text{bat},i}, I_{\text{bat},i}, P_{b,i}, P_{c,i}]^{\top}$, $u_i = [P_{\text{grid},i}, Q_{c,i}]^{\top}$, and $v_i = Q_{c,i}$. The disturbance ${d}_{i}$ will be discussed later in Section~\ref{sec:sec_4c}.

\section{Ensemble Nonlinear Model Predictive Control Based Economic Dispatch for Home Energy Management} \label{sec:sec_4}

\subsection{Operating Costs} \label{sec:sec_4a}

As presented in (\ref{eq:eq_1a}), the control objective is to minimize the accumulated operating cost over the optimization horizon $T$. Particularly, the control objective considers 1) to reduce the costs of electricity transactions or net trading costs and 2) to reduce the battery usage cost due to its degradation. Other objectives that can be incorporated in future studies include the emission level, user satisfaction cost, response fatigue, the penalty due to the dispatch schedule mismatch, and the incentive of participating in grid regulation such as frequency and voltage control \cite{7984899}. In the present work, the trading cost per unit time is expressed as
\begin{subequations} \label{eq:eq_13}
    \begin{equation} \label{eq:eq_13a}
    {\text{{Cost}}_{i}^{\text{trade}}}({{u}_{i}})=-{{\lambda }_{i}} {{u}_{i}}=-{{\lambda}_{i}} {{P}_{\text{grid},i}}   \quad [\$/\text{h}]
    \end{equation}

On the other hand, the cost incurred by the battery degradation is
\begin{equation}  \label{eq:eq_13b}
    {\text{Cost}_{i}^{\text{bat}}} = {\kappa} ({\Delta {Q}_{\text{loss},i}}/{\Delta t})   \quad [\$/\text{h}]
\end{equation}

The capacity fade rate in \eqref{eq:eq_13b} is obtained using \eqref{eq:eq_2a}--\eqref{eq:eq_2c} and \eqref{eq:eq_5} as
\begin{equation}  \label{eq:eq_13c}
   \frac{\Delta {Q}_{\text{loss},i}}{\Delta t} = \frac{{Q}_{\text{loss},i+1} - {Q}_{\text{loss},i}}{\Delta t} = -I_{\text{sr},i}   \quad [\text{Ah}/\text{h}]
\end{equation}

The coefficient ${\kappa}$ in \eqref{eq:eq_13b} converts the capacity fade rate into an economic loss rate, given by
    \begin{equation} \label{eq:eq_13d}
    {\kappa}=\frac{N_{\text{cell}}{\xi_{\text{cell}}}}{(1-K\%){{Q}_{0}}}   \quad [\$/\text{Ah}].
    \end{equation}
where $\xi_\text{cell}$ is the overall investment per battery cell, and it is assumed that the end-of-life of the battery is reached when capacity has dropped to $K\%$ of its pristine capacity $Q_{0}$. The total cost per unit time (running cost) is thus
    \begin{equation} \label{eq:eq_13e}
    {\text{Cost}_{i}^{\text{op}}}={\text{Cost}_{i}^{\text{trade}}}+  {\text{Cost}_{i}^{\text{bat}}}  \quad [\$/\text{h}].
    \end{equation}
\end{subequations}

\subsection{Operating Constraints} \label{sec:sec_4b}

For \eqref{eq:eq_1e}--\eqref{eq:eq_1g}, the following inequality constraints must be satisfied to ensure safe and reliable operation:
\begin{subequations} \label{eq:eq_14}
    \begin{align}
    {\text{SoC}}_{\text{min}} &\le \text{SoC}_{i}\le {\text{SoC}}_{\text{max}} \label{eq:eq_14a}\\
    {{V}}_{\text{min}}&\le {{V}_{\text{bat},i}}\le {{V}}_{\text{max}} \label{eq:eq_14b}\\
    -{{I}}_{\text{max}}&\le {{I}_{\text{bat},i}}\le {{I}}_{\text{max}} \label{eq:eq_14c}\\
    -{{P}_{b,\text{max}}}&\le {{P}_{b,i}}\le {{P}_{b,\text{max}}} \label{eq:eq_14d}\\
    -{{P}_{\text{grid},\text{max}}}&\le {{P}_{\text{grid},i}}\le {{P}_{\text{grid},\text{max}}}  \label{eq:eq_14e}\\
   {{T}}_{\text{min}}&\le {{T}_{{\text{bat}},i}}\le {{T}}_{\text{max}} \label{eq:eq_14f} \\
    -{{P}}_{c,\text{max}}&\le {{P}_{c,i}}\le {{P}}_{c,\text{max}} \label{eq:eq_14g}
    \end{align}
$\forall i \in \mathcal{S}_T$, where \eqref{eq:eq_14a} defines the expected normal operating range of the Li-ion batteries to avoid overcharge and over-discharge, \eqref{eq:eq_14b}--\eqref{eq:eq_14d} are determined by the voltage, current, and power capability of the battery system converter, respectively, \eqref{eq:eq_14e} specifies the power rating of the interfacing transformer, and inequalities \eqref{eq:eq_14f} and \eqref{eq:eq_14g} specify the temperature range and power consumption for battery temperature control, respectively.
\end{subequations}

\subsection{Forecast of Disturbance} \label{sec:sec_4c}

The disturbances, including the output power $P_{\text{PV}}$ of the PV panels, residential load $P_{\text{load}}$, real-time electricity price $\lambda$, and ambient temperature $T_{\text{amb}}$, are required to be forecasted for the purpose of predictive dispatch. While we denote the actual disturbance vector as ${d}_i = [{P}_{\text{PV},i}, {P}_{\text{load},i}, {\lambda}_{i}, T_{\text{amb},i}]^{\top}$, the forecasted disturbance vector is expressed by $\hat{d}_i = [\hat{P}_{\text{PV},i}, \hat{P}_{\text{load},i}, \hat{\lambda}_{i}, \hat{T}_{\text{amb},i}]^{\top} = {d}_i + {w}_i$, where $w_i = [w_{\text{PV},i}, w_{\text{load},i}, w_{\lambda,i}, w_{T\text{amb},i}]^{\top}$ is the forecast error.

Although there are numerous techniques for predicting the disturbances in the literature, such as solar generation in \cite{ahmedr2020rser} and residential load in \cite{wenl2020epsr}, high-performance forecasts always require massive historical data and computationally-demanding statistical/machine learning algorithms. However, the data storage and computation capabilities of an individual residential prosumer may be very limited. Hence, the level of uncertainty and stochasticity in the forecast can vary significantly depending on the practical configuration of HEMSs. Since the proposed dispatch strategy should be robust and general enough, the development of improved forecast algorithms will not be focused in the present investigation. We assume only limited local capacity is available to store the recent historical forecast errors (e.g., several days) for an individual household, while the future forecasts on PV power, load, electricity prices, and ambient temperature are provided by the DSO and can be accessed via smart meters.

\subsection{EnNMPC-Based Economic Dispatch} \label{sec:sec_4d}

The nonlinear optimization problem \eqref{eq:eq_1} is difficult to solve due to its long prediction horizon, large number of decision variables, the presence of various uncertainties in the forecast, and the delayed control nature. We propose a tractable EnNMPC-based solution with a two-stage strategy as follows.

The first stage is to predict the state at the end of the delay horizon $D$, denoted by $\hat{x}_{k+n_D}$. To do so, the control sequence $\mathbf{u}_{k:k+n_D - 1}(\cdot)$ is first calculated with the pre-determined
dispatch schedule
and considering the constraint \eqref{eq:eq_1h}. One can then use $\mathbf{u}_{k:k+n_D -  1}(\cdot)$ to predict the future state $\hat{x}_{k+n_D}$ by forwarding (\ref{eq:eq_1c}), with the estimated initial state $\hat{x}_{k}$ and the forecasted disturbance sequence $\mathbf{\hat{d}}_{k:k+n_D-1}(\cdot) = [\hat{d}_{k},\cdots, \hat{d}_{k+n_D-1}]$.

To address the uncertainty in the forecast, $m$ realizations $\hat{d}_{i}^{(j)}  \in \mathbb{R}^{n_d} , \forall j \in \mathcal{J} := \{1, 2, \cdots, m\}$ (namely, a \emph{disturbance ensemble}) of the disturbance forecast $\hat{d}_{i}$ over the delay horizon $D$ are generated and used, i.e.,
\begin{subequations} \label{eq:eq_15}
\begin{align}
\hat{x}_{k}^{(j)}  &= \hat{x}_k \label{eq:eq_15a}, \quad \forall j \in \mathcal{J} \\
\hat{x}_{i + 1}^{(j)} &= f(\hat{x}_{i}^{(j)}, u_{i} ,\hat{d}_{i}^{(j)}), \quad \forall i \in \mathcal{S}_D, \forall j \in \mathcal{J} \label{eq:eq_15b}
\end{align}
The realization $\hat{d}_{i}^{(j)}$ is named an \emph{ensemble member} of the disturbance ensemble defined at time instant $i$.

The predicted state $\hat{x}_{k+n_D}$ is calculated with weighted ensemble members $\hat{x}_{k+n_D}^{(j)}$. Considering that the delay horizon is usually short, \eqref{eq:eq_15b} can be approximated linearly with sufficient accuracy. Furthermore, under the assumption that $d_i^{(j)}$ is normally distributed, it can be readily shown $\hat{x}_{k+n_D}^{(j)}$ are normally distributed. In this condition, the predicted state at the end of $D$ can be approximated by the \emph{ensemble mean}, i.e.,
\begin{equation}
\hat{x}_{k+n_D}  = \frac{1}{m} \sum_{j \in \mathcal{J}} \hat{x}_{k+n_D}^{(j)}.
\end{equation}

\end{subequations}

The second stage is to determine ${{{\mathbf{P}}_{\text{grid},k}}(\cdot)}$ by solving a robust optimization problem based on the predicted state $\hat{x}_{k+n_D}$ from the first stage, an ensemble of $m$ realizations of the disturbance  ${\hat{d}}_i^{(j)}, i \in \mathcal{S}_D, \forall j \in \mathcal{J}$, and the constraints (\ref{eq:eq_1c})--(\ref{eq:eq_1h}) placed on the decision variables and the ensemble members of the state and output variables. Instead of determining all $v_i = Q_{c,i}, i \in \mathcal{S}_T$ as in \eqref{eq:eq_1}, we align the temperature control to the dispatch interval to reduce the number of decision variables. Specifically, we define ${{{\mathbf{Q}}_{c,k}}(\cdot)} := [Q_{c,k + n_D}, Q_{c,k + n_D + n_\text{DI}},  \cdots, Q_{c,k + n_D + (\frac{T}{{{\text{DI}}}}-1) n_\text{DI}}] \in \mathbb{R}^\frac{T}{{\text{DI}}}$, and  \eqref{eq:eq_1}  will be replaced by
\begin{subequations} \label{eq:eq_16}
    \begin{align}
    \min_{\mathbf{P}_{\text{grid},k}(\cdot),\mathbf{Q}_{c,k}(\cdot)} \quad &\sum\limits_{j \in \mathcal{J}}{\sum\limits_{i \in \mathcal{S}_T}{{\text{Cost}_{i}^{\text{op},(j)}}}} \label{eq:eq_16a} \\
\textrm{s.t.} \quad & x_{k+n_D}^{(j)} = \hat{x}_{k+n_D} \label{eq:eq_16b}\\
&x_{i + 1}^{(j)} = f(x_{i}^{(j)}, u_{i} ,\hat{d}_{i}^{(j)}) \label{eq:eq_16c}\\
   & x_{i + 1}^{(j)} \in \mathcal{X} \subset {{\mathbb{R}}^{n_x}} \\
   & y_{i }^{(j)} \in \mathcal{Y} \subset {{\mathbb{R}}^{n_y}} \\
   & {{u}_{i}} \in \mathcal{U} \subset \mathbb{R} \\
   & [P_{\text{grid},i}, Q_{c,i}]^{\top} = \left.u_{l}\right|_{l =\lfloor {i}/{n_\text{DI}} \rfloor n_\text{DI}}
\end{align}
\end{subequations}
where $\forall i \in \mathcal{S}_T, \forall j \in \mathcal{J}$.  The condition \eqref{eq:eq_16b} is called the non-anticipativity constraint and corresponds to a robust horizon of one \cite{marti2015cce,garcia2019acc}. Hence, we do not consider the feedback in the prediction which means the future control actions will not affect the uncertainty in the disturbance. Note that the above algorithm is repeated every $n_M$ sampling intervals as time proceeds, and thus only the first $M$ elements in the dispatch schedule ${\mathbf{P}_{\text{grid},k}(\cdot)}$ are the firm commitment.

It is worth noting that the formulation of the optimization problem \eqref{eq:eq_16} is very general and can be extended to some existing scenarios for HEMS. For example, if the delay horizon $D$ is set to 24 hours and the modification horizon $M$ is zero, the problem is equivalent to the conventional day-ahead home energy dispatch \cite{dinh2022scirep}. In any case, the proposed battery model and the EnNMPC algorithm can still be incorporated into the new optimization model.

\subsection{Ensemble Generation} \label{sec:sec_4e}

Note that in \eqref{eq:eq_15b} and \eqref{eq:eq_16c}, the disturbance ensemble ${\hat{d}_{i}^{(j)}}$ is needed for $i \in \mathcal{S}_D$ and $i \in \mathcal{S}_T$, respectively. A general method to generate the disturbance ensemble is proposed in this subsection. The disturbance ensemble over the prediction horizon is given by
\begin{equation} \label{eq:eq_17}
\hat{d}_i^{(j)} = \hat{d}_i + w_i^{(j)} \quad \forall i \in \mathcal{S}_P, \forall j \in \mathcal{J}
\end{equation}
where $\hat{d}_i$ is obtained day-ahead with a specific forecast technique and $w_i^{(j)} \in \mathbb{R}^{n_d}$ is a realization of the forecast error $w_i = \hat{d}_i - d_i$.

The sequence $\{w_i\}$ is considered a multivariate random process. Intuitively, the forecast error for HEMS-related variables can have non-zero mean but obvious periodic characteristics. As observed in existing works such as \cite{VAGROPOULOS20169}, forecast error in PV power is found to be normally distributed based on hourly resolution. However, when data resolution increases, the time series of the forecast errors are nonstationary and cannot be regarded as white noise. In this condition, we independently generate the \emph{forecast error ensemble} at time instant $i$ by considering the statistical parameters $\Theta_i = \{{\theta}_{1,i}, {\theta}_{2,i}, \cdots, {\theta}_{p,i}\}$, i.e., the first to $p$th moments. We denoted a general distribution as
\begin{equation}
 w_i^{(j)} \sim \mathcal{D}({\theta}_{1,i}, {\theta}_{2,i}, \cdots, {\theta}_{p,i}) \label{eq:eq_18}
\end{equation}

 A three-step method to learn the parameter set is proposed as follows.

1) Divide the time of a day into $S$ segments. For the forecast of disturbance error at a future time instant $i \in \mathcal{S}_P$, we denote the segment number as $s_i \in \{1,2, \cdots, S\}$.

2) Select the historical forecast errors with the same segment number $s_i$ in the previous $H$ days. We denote the corresponding historical forecast errors as $w_n, n \in \mathcal{S}_{H,i}$

3) Approximate the parameters in $\Theta_i$ with sampled values. The first and the second moments can be readily approximated by sample mean and the unbiased sample covariance, i.e.,
\begin{align}
{\theta}_{1,i} = \mu_{i}  &= \frac{1}{n_H} \sum_{n \in \mathcal{S}_{H,i}} w_{n}  \label{eq:eq_19} \\
{\theta}_{2,i} = \Sigma_{i} &= \frac{1}{n_H-1} \sum_{n \in \mathcal{S}_{H,i}}( {{w}}_n - {{\mu}}_i)( {{w}}_n - {{\mu}}_i)^{\top} \label{eq:eq_20}
\end{align}
where $\mu_i\in \mathbb{R}^{n_d}$ is the mean vector,  ${\Sigma}_i \in \mathbb{R}^{n_d \times n_d} $ is the covariance matrix, and $n_H$ is the size of $\mathcal{S}_{H,i}$.
The multivariate higher moments such as the third moment (skewness) and the fourth moment (kurtosis) can also be approximated using sampled values, although the calculations are more complex. The reader can refer to various works such as \cite{mardia1970} for detailed discussion.

\begin{figure}[!bt]
\centering
  \includegraphics[width=0.48\textwidth]{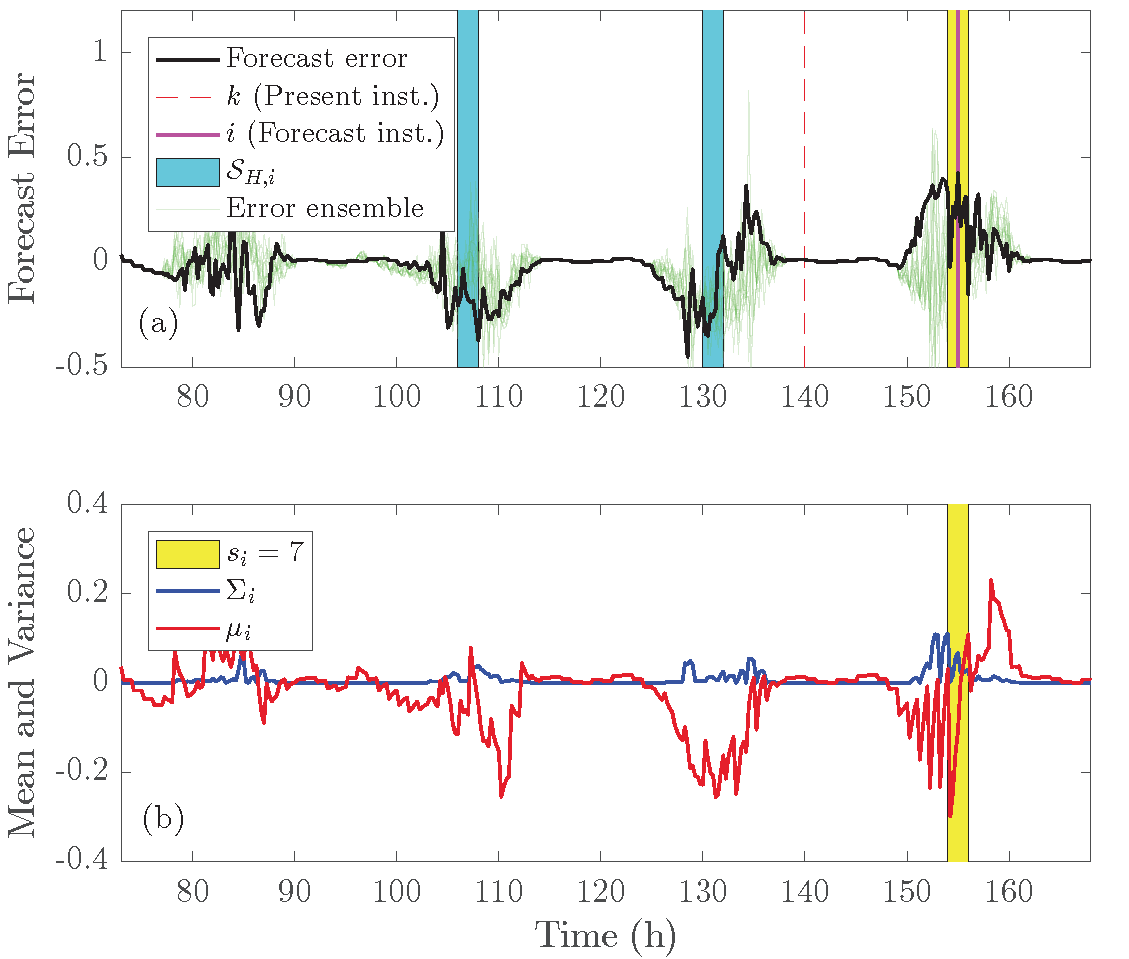}
  \caption{Example of ensemble generation procedure ($m = 10$): (a) Forecast error and error ensemble, and (b) estimated mean and variance. }
  \label{fig:fig_3}
\end{figure}

Following the above steps, one can calculate the parameter set $\Theta_i, \forall i \in \mathcal{S}_P$, with which the distribution $\mathcal{D}$ can be identified for generating the error ensemble according to \eqref{eq:eq_14}. In fact, the first $n_P - n_M$ parameter sets, i.e., $\Theta_i, i \in \{k , \cdots, k + n_P - n_M - 1\}$, have been obtained in previous steps. Hence, only the last $n_M$ parameter sets ($i \in \{k + n_P - n_M - 2, \cdots, k + n_P - 1\}$) need to be estimated at each time instant $k$.

Fig.~\ref{fig:fig_3} shows an example of how the disturbance ensemble is generated based on the proposed method with $H = 2$ (days), $S = 12$ (segments per day), and ensemble size $m = 10$. For demonstration purposes, the forecast error is one-dimensional, as shown in Fig.~\ref{fig:fig_3}(a). The forecast errors are calculated with the real-world prosumer PV power data and a machine learning technique as described in the next section. In this example, we only use the first and the second moments to describe the statistical characteristics. Therefore, we select $p = 2$ in \eqref{eq:eq_18} and thus $w_i^{(j)} \sim \mathcal{N}(\mu_i, \Sigma_i)$ is normally distributed. The present time instant is at the beginning of $140$ h, and thus only the historical forecast errors earlier than $140$ h are available. The target forecast time instant of $155$ h is in the segment $s_i = 7$. According to the proposed algorithm, the forecast error data in segment $7$ of the previous two days, indicated by the shaded areas in Fig.~\ref{fig:fig_3}(a), are used to calculate the mean and variance according to  \eqref{eq:eq_19} and \eqref{eq:eq_20}, for the interval between $154$ h and $156$ h as indicated in Fig.~\ref{fig:fig_4}(b). The corresponding ensemble of forecast error generated for the interval between $154$ h and $156$ h as shown in Fig.~\ref{fig:fig_3}(a) is compared with the true error measured in later steps.

\section{Illustrative Examples} \label{sec:sec_5}

\subsection{Experimental Setup and Model Comparison}

In this section, the efficacy of the proposed EnNMPC algorithm will be validated by means of computer simulation with real-world prosumer data. All algorithms were implemented in MATLAB R2016a environment on a PC with a processor @ 2.6 GHz and 8GB RAM.

\begin{figure}[!tb]
\centering
  \includegraphics[width=0.48\textwidth]{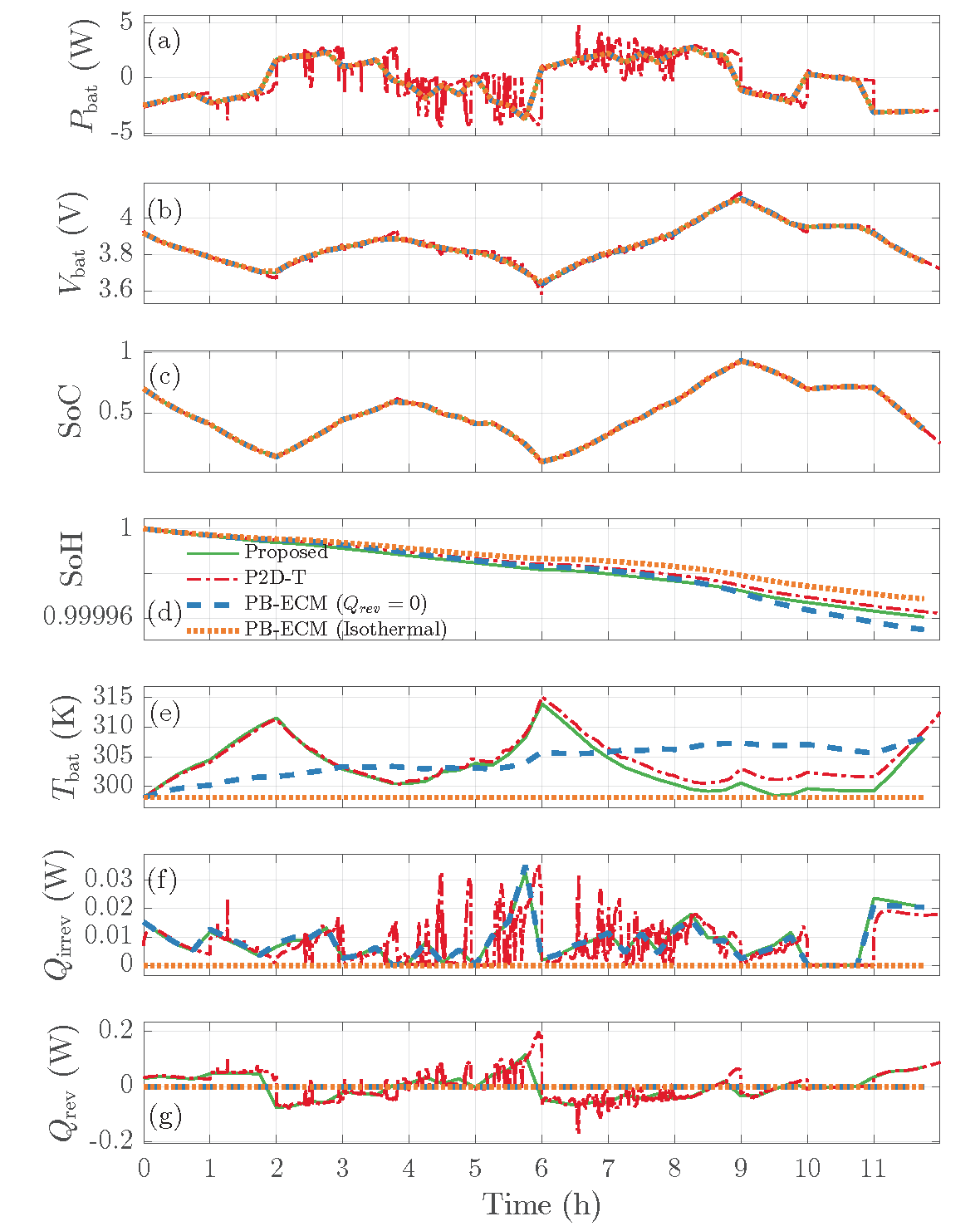}
  \caption{Comparison of battery models:  (a) battery power, (b) battery voltage,  (c) SoC, (d) SoH, (e) battery temperature, (f) irreversible heat generation, and (g) reversible heat generation.}
  \label{fig:fig_4}
\end{figure}

We first validate the proposed temperature-enhanced PB-ECM of the Li-ion battery by comparing it with an experimentally verified partial differential equation-based pseudo-two-dimensional (P2D) model with thermal dynamics (P2D-T) in Fig.~\ref{fig:fig_4}. The P2D-T model was simulated using the LIONSIMBA package from where most model parameters are obtained (as shown in Table I in Appendix). The detail of this high-fidelity package is described in \cite{torchio2016jes}. Note that the battery model parameters can be identified by the battery management system \cite{MIGUEL2021103388}. Given that the influence of battery degradation on model parameters is very small in the short term, these parameters can be considered unchanging within an optimization horizon and updated on a regular basis, such as weekly or monthly.  In addition, in Fig.~\ref{fig:fig_4} we also compare the proposed model with an isothermal PB-ECM and a PB-ECM that only considers the irreversible heat generation. The battery current is calculated based on measured 1-s solar irradiance (Global Horizontal AP4 \cite{nrelsolardata}) with a randomly generated dispatch schedule. For the PB-ECM models, the sampling time is 15 min, while the P2D-T model requires a sample time of 1 s for solving. Although some fast dynamics due to rapidly changing PV power have been ignored, the proposed model can accurately predict the variables that are important for HEMS studies, such as SoC, voltage, temperature, SoH, etc.
Fig.~\ref{fig:fig_4}(e) also shows that the temperature may vary significantly under dynamic operating conditions. Both high and low temperatures can significantly affect battery safety and lifetime, especially for HEMS in a cold or hot environment. The temperature dynamics are not correctly reproduced with the isothermal PB-ECM and the PB-ECM that ignores the reversible heat, which can then cause inaccurate prediction of degradation, as illustrated in Fig.~\ref{fig:fig_4}(c). Hence, incorporating the battery temperature model with reversible heat generation and the corresponding control into the HEMS strategy is deemed advantageous for the present applications.

Next, one-year data recorded from March 1, 2016, to February 28, 2017, were used to validate the proposed EnNMPC scheme. Specifically, the PV generation, load profile, and ambient temperature for the residential users were measured in 15-minute resolution \cite{opsd2020household}, while the half-hourly updated real-time electricity prices were assumed for the residential prosumer \cite{ameo}. The electricity prices are increased by a factor of five to consider a highly profitable scenario for the prosumers. The day-ahead forecasts of these disturbances were obtained using an artificial neural network where input data include the season, month of the year, hour for the day, and the measured disturbances in the previous day.
We assume that the initial battery states $\text{SoC}(0) = 0.5$ and $\text{SoH}(0) = 1$ are known, and the initial charges $Q_1^+(0)$ and $Q_1^-(0)$ are calculated according to \eqref{eq:eq_7} and \eqref{eq:eq_8}. It is also assumed that the present battery state $\hat{x}_k$ has been accurately monitored with a well-designed battery state estimator, by which model uncertainty and measurement errors can be taken into consideration \cite{liyang2021tie}.

The finite-time optimal control problem \eqref{eq:eq_16} was reformulated as a constrained nonlinear programming problem and solved  with sequential quadratic programming using MATLAB Optimization Toolbox function \texttt{fmincon}. Due to high model nonlinearity and the inevitable disturbance errors, to improve the recursive feasibility when solving the optimization problem, soft constraints are applied on battery SoC and temperature during the online operation. Specifically, we modified the constraints on SoC and temperature as per:
\begin{align}
\text{So}{{\text{C}}_{\text{min}}}-{{\epsilon }_{1}} &\le \text{SoC}_{i+1}^{(j)}\le \text{So}{{\text{C}}_{\text{max}}}+{{\epsilon }_{2}} \\
{{T}_{\text{min}}}-{{\epsilon}_{3}} &\le T_{\text{bat},i+1}^{(j)}\le {{T}_{\text{max} }}+{{\epsilon }_{4}}\\
 &\epsilon_1,  \epsilon_2, \epsilon_3, \epsilon_4 \ge 0
\end{align}
where $\epsilon_1,  \epsilon_2, \epsilon_3, \epsilon_4$ are four new decision variables. Furthermore, an additional term was added to the cost function \eqref{eq:eq_16a}:
\begin{align}\left[ \begin{matrix}
   {{\epsilon }_{1}} & {{\epsilon }_{3}}  \\
\end{matrix} \right]{{W}_{1}}{{\left[ \begin{matrix}
   {{\epsilon }_{1}} & {{\epsilon }_{3}}  \\
\end{matrix} \right]}^{\top}}+\left[ \begin{matrix}
   {{\epsilon }_{2}} & {{\epsilon }_{4}}  \\
\end{matrix} \right]{{W}_{2}}{{\left[ \begin{matrix}
   {{\epsilon }_{2}} & {{\epsilon }_{4}}  \\
\end{matrix} \right]}^{\top}}\end{align}
where $W_1$ and $W_2$ are two positive-definite weight matrices.

The dispatch rules stipulated by the DSO are: $\text{DI} = 1$~h, $D = 2$~h, $T = 12$~h,  $P = 14$~h, and $M = 1$~h. The operating constraints are given as ${P}_{b,\text{max}} = 12$ kW, $P_{\text{grid},\text{max}} = 10$ kW, ${P}_{c,\text{max}} = 0.1$ kW,  ${I}_{\text{max}} = 0.873$~A ($0.5$C-Rate), ${V}_{\text{max}} = 4.2$ V,  ${V}_{\text{min}} = 3.2$ V, ${\text{SoC}}_{\text{max}} = 0.99$, ${\text{SoC}}_{\text{min}} = 0.01$, $T_{\text{min}} = 296.15~\text{K}~(23~^{\circ}\text{C})$, and $T_{\text{max}}= 300.15~\text{K}~(27~^{\circ}\text{C})$. In addition, the battery cell capacity is $Q_0 = 1.747$ Ah, $N_{\text{cell}} = 4000$, $K\% = 60\%$, $\xi = \$1$, $\eta_c = 700\%$, $C_T = 0.0278~\text{kWh/K}$, $R_T = 200~\text{K/kW}$. The nonlinear functions and other parameters in the battery model are given in the Appendix.

\subsection{Results and Discussion}

\begin{figure}[!bt]
\centering
  \includegraphics[width=0.48\textwidth]{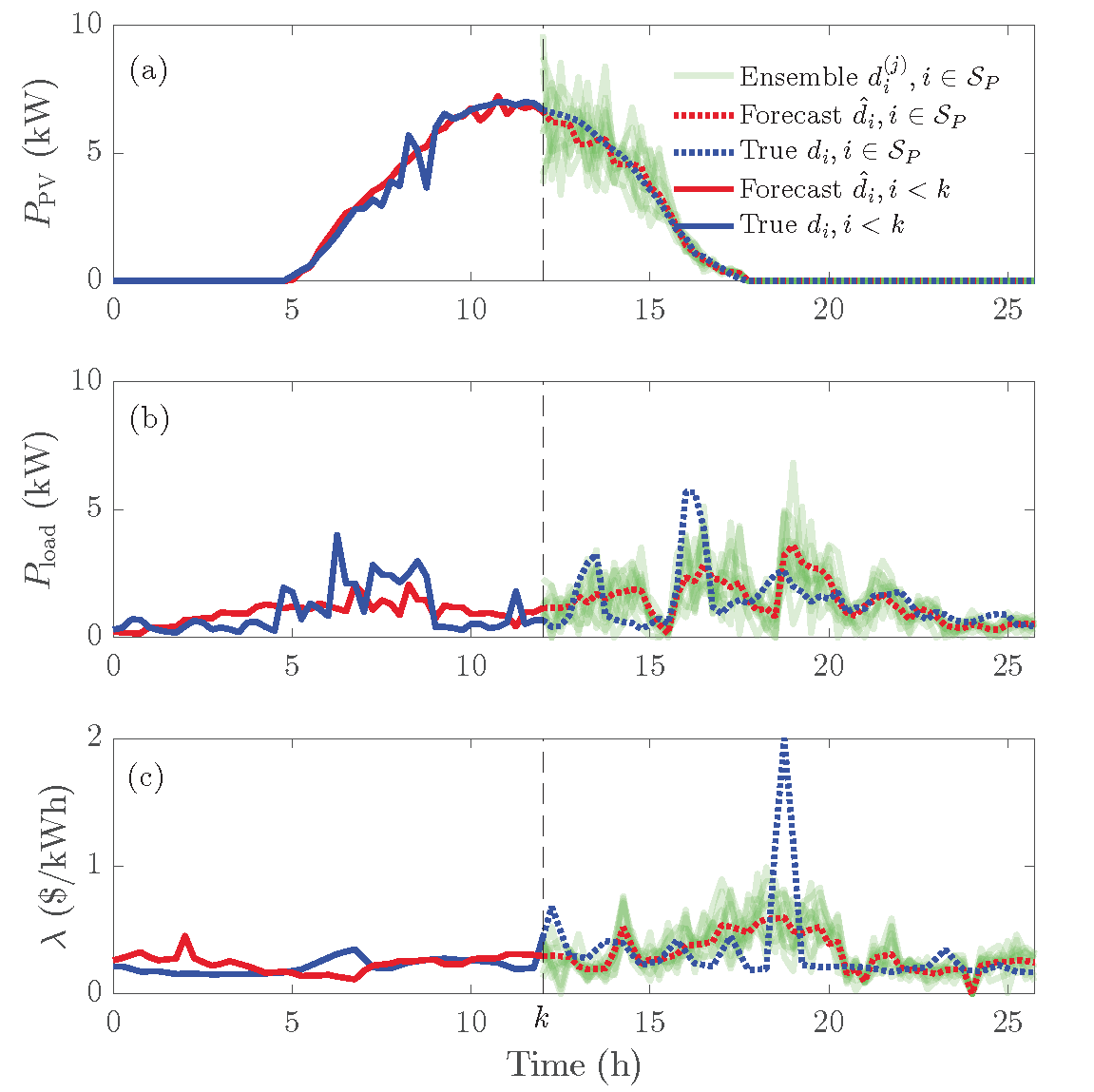}
  \caption{Example of 24-h forecast and generated ensemble ($m = 10$): (a) PV power, (b) load, and (c) electricity price. }
  \label{fig:fig_5}
\end{figure}
\begin{figure}[!bt]
\centering
  \includegraphics[width=0.48\textwidth]{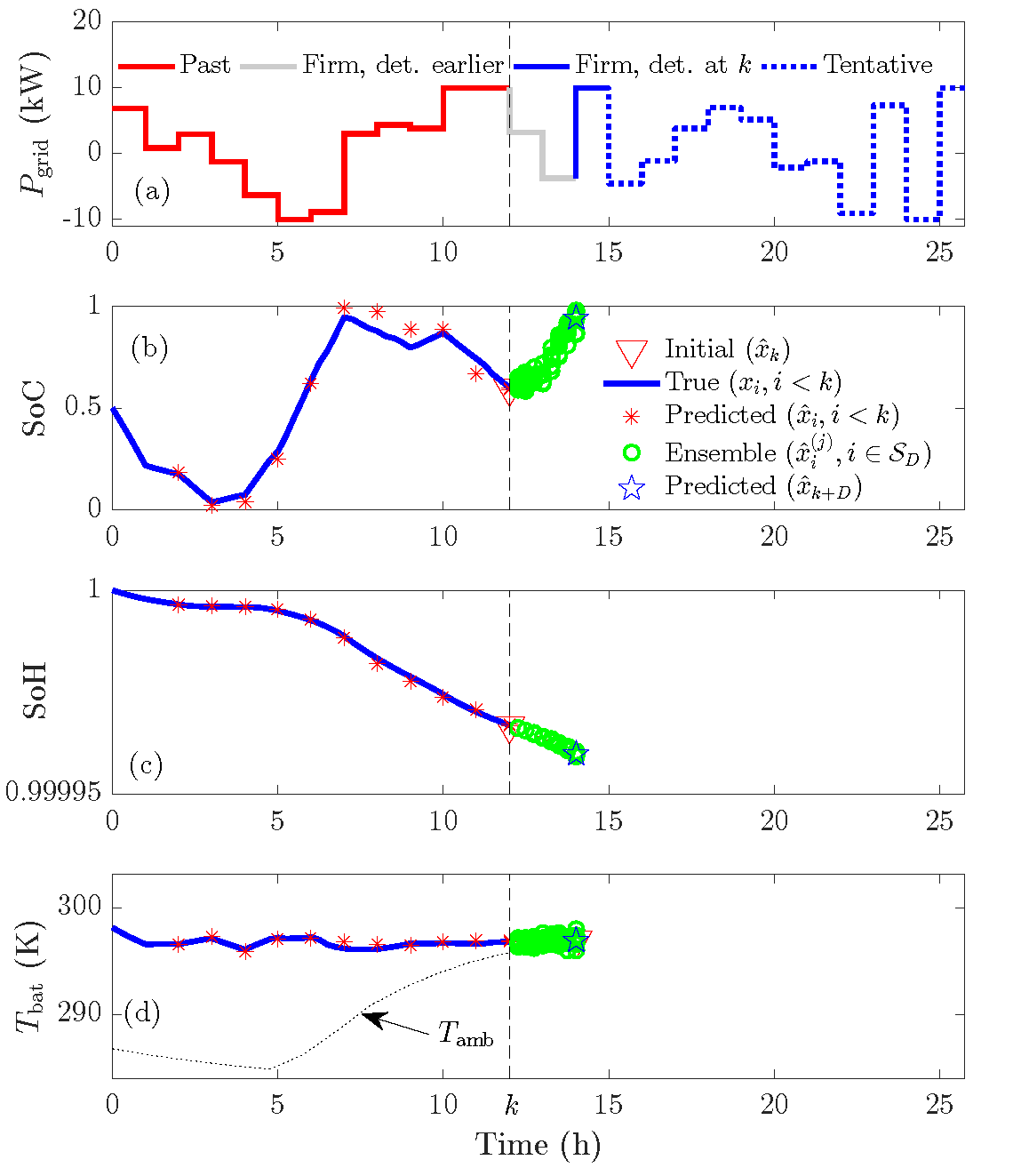}
  \caption{Example of 24-h forecast and generated ensemble ($m = 10$): (a) dispatch schedule, (b) SoC, (c) SoH, and (d) battery temperature.}
  \label{fig:fig_6}
\end{figure}

An example of the simulated results based on the proposed algorithm is shown in Fig.~\ref{fig:fig_5} and Fig.~\ref{fig:fig_6} where the present time is at the beginning of $12$ h. At this time instant, with the proposed method described in Section~\ref{sec:sec_4e}, the disturbance ensembles are generated and shown in Fig.~\ref{fig:fig_5} along with the forecast and the measured disturbance. It can be seen that although there are significant forecast errors, the generated ensemble can cover a wide range of trajectories of the measured disturbances. Fig.~\ref{fig:fig_6}(a) presents the past, firmly committed, and tentative (adjustable) dispatch schedule as observed at the present time instant $k = 12$. The $2$-h schedule between $12$~h and $14$~h is determined before $12$~h and has to be applied in the subsequent two hours. The $1$-h schedule between $14$~h and $15$~h is newly determined via solving the optimization problem and is not subjected to any future change. The schedule after $15$~h, however, can be revised and resubmitted later. In Fig.~\ref{fig:fig_6}(b)--Fig.~\ref{fig:fig_6}(d), the initial value, measured, predicted, and the ensemble of battery SoC, SoH, and temperature are shown. They are back-calculated based on the respective state $x$ according to \eqref{eq:eq_7} and \eqref{eq:eq_8}. Clearly, the predicted states are accurate despite significant forecast errors.

It should be noted that the generated ensemble based on the proposed method may lead to physically unrealistic realizations. For example, the generated PV power based on the present forecast and its learned statistics might exceed the maximum possible PV power. We found such a strategy does not affect the effectiveness of the algorithm, but it can improve the feasibility of the optimization. In-depth investigations of the impact of unrealistic realizations will be focused in our future works.

\subsection{Selection of the Ensemble Size}

\begin{figure}[!bt]
\centering
  \includegraphics[width=0.48\textwidth]{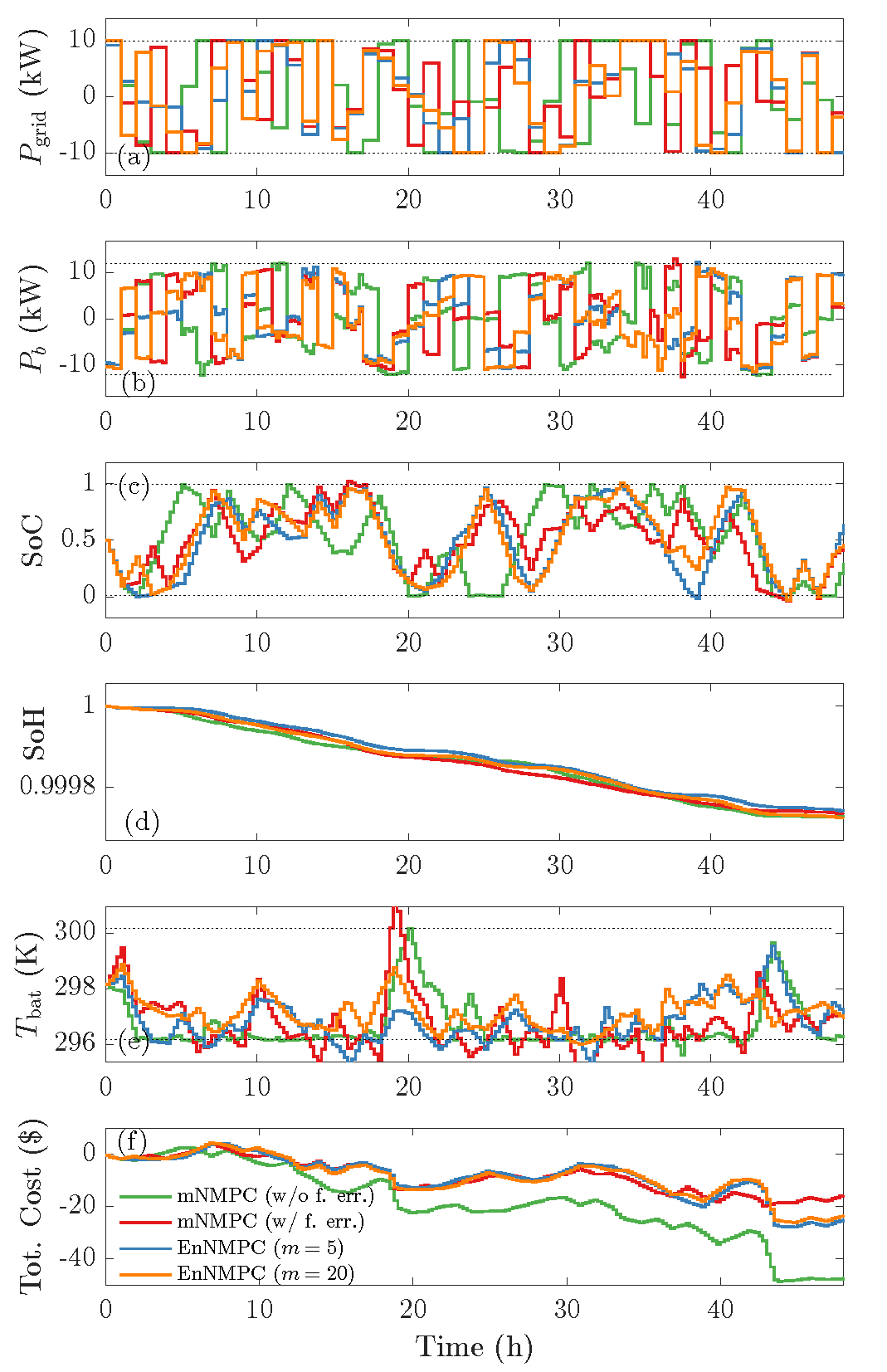}
  \caption{Example of 48-h simulation results of (a) dispatch schedule, (b) battery power, (c) SoC, (d) SoH, (e) temperature, and (f) accumulated total operating costs.}
  \label{fig:fig_7}
\end{figure}

\begin{figure}[!bt]
\centering
  \includegraphics[width=0.48\textwidth]{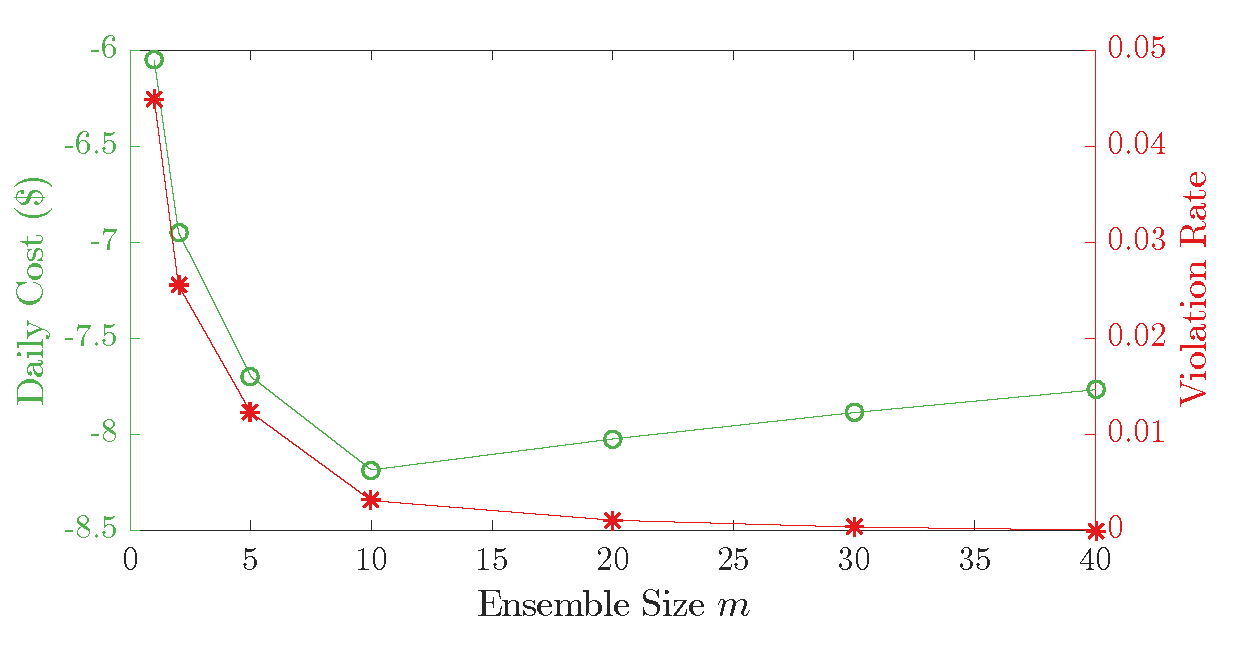}
  \caption{Relationship between the ensemble size, average daily cost, and violation rate based on one-year simulation results.}
  \label{fig:fig_8}
\end{figure}

\begin{figure}[!bt]
\centering
  \includegraphics[width=0.48\textwidth]{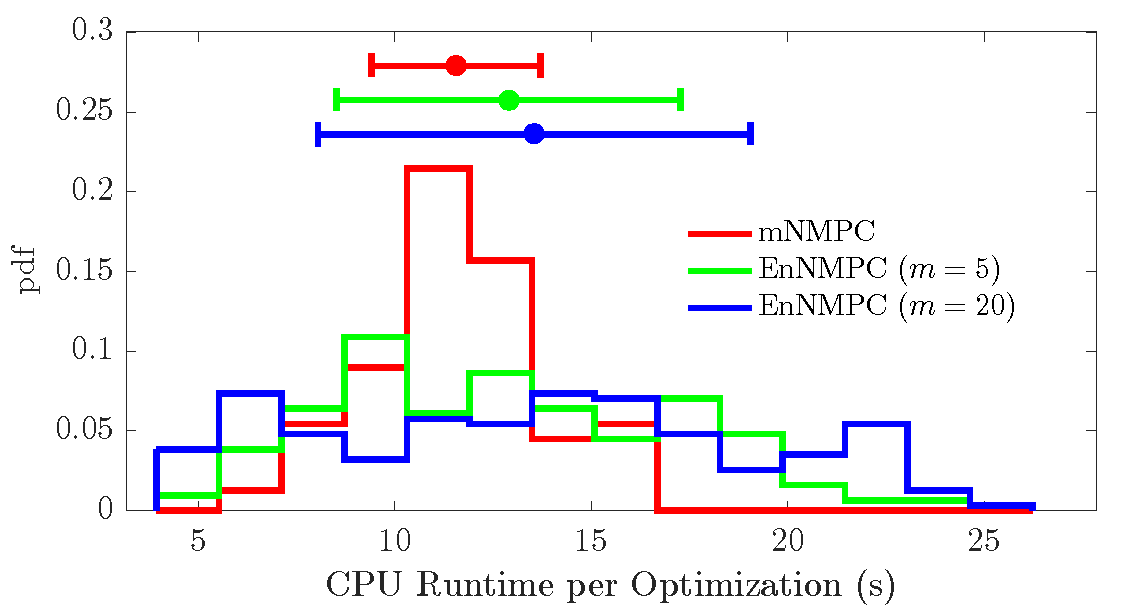}
  \caption{Comparison of CPU runtime histogram of different algorithms.}
  \label{fig:fig_9}
\end{figure}

The main tuning parameter of the proposed algorithm is the ensemble size $m$. In order to investigate the effect of different $m$ on the HEMS performance, a series of simulations were carried out. An example of the effect of different $m$ on the first 48-h simulation results is shown in Fig.~\ref{fig:fig_7} with the ensemble size $m$ being $5$ and $20$, respectively. The results are compared with that obtained using the mNMPC-based dispatch strategy as presented in \mbox{\cite{liyang2018iecon}} which is in fact equivalent to $m = 1$ in the proposed EnNMPC strategy. Furthermore, a baseline method is also compared, where the optimization problem is solved using an ideal mNMPC without forecast errors. Whilst this baseline result is not realistic and constitutes an unachievable benchmark, it is included here for comparison purposes.

The dispatched power, battery power, battery SoC, SoH, and temperature are shown in Fig.~\ref{fig:fig_7}(a) to Fig.~\ref{fig:fig_7}(e), respectively. It can be observed from Fig.~\ref{fig:fig_7}(c) and Fig.~\ref{fig:fig_7}(e) that the battery attempts to operate under low-SoC and low-temperature regions to reduce the aging rate. This is determined by the battery characteristics: the SEI film growth tends to be slower at lower SoC and lower temperatures. However, unlike the unrealistic benchmark solution where prior disturbance error is exactly known, due to the uncertainty in the forecast, the SoC and the battery temperature shown in Fig.~\ref{fig:fig_7}(c) and Fig.~\ref{fig:fig_7}(e) are observed to violate the constraints at times. It can be observed that the chance of the violation of constraints can be reduced by increasing the ensemble size. However, the increased ensemble size requires additional computational resources. More importantly, as seen in Fig.~\ref{fig:fig_7}(f), a large ensemble size may lead to a conservative dispatch strategy with increased operating costs. Furthermore, we do not see a consistent relationship between the total costs and the ensemble size $m$ in Fig.~\ref{fig:fig_7}(f) and this short time period is insufficient for reliable cost evaluation.

Hence, long-term simulations over a year's data were conducted and the results are summarized. We obtain the relationships between the ensemble size and some statistical indices including 1) the daily costs, calculated by summing up all $\text{Cost}_{i}^{\text{op}}$ in a day, 2) the constraint violation rate, defined as the number of time instants with constraint violation divided by the total time instants,
as shown in Fig.~\ref{fig:fig_8}. It is interesting to observe that in this case, the daily cost is always negative, indicating that the prosumer can always receive net income. Furthermore, an optimal ensemble size exists at $m = 10$ where the daily income is high as $\$8.2$: The prosumer can earn 36\% more than the case with $m = 1$ (mNMPC-based) where the daily income is $\$6.03$. When $m > 10$, the additional benefit from reducing the constraint violation rate is marginal and the corresponding strategy can be more conservative, leading to a slow and continuing increase in the average daily operating costs.

Finally, the CPU runtimes per optimization for different algorithms are compared in Fig.~\ref{fig:fig_9} in terms of histogram and boxplot. Surprisingly, it is observed that by increasing the size of ensemble, the mean computational runtime per optimization is not much affected, while the variance increases more significantly. This result shows an advantage of the proposed method over many stochastic and robust programming and mNMPC schemes commonly found in the literature.

\section{Conclusions}  \label{sec:sec_6}
A novel minimum-cost short-term dispatch strategy is proposed for residential energy systems that feature photovoltaic panels and Li-ion batteries. The system is assumed to operate under a dynamic environment of the distribution energy market. The optimal dispatch power at the point of connection is determined and scheduled using a novel strategy based on ensemble nonlinear model predictive control (EnNMPC), with the objective to reduce the overall operating cost of the residential units. In the proposed strategy, a physics-based equivalent circuit model is developed to predict the dynamical behaviors of Li-ion batteries, which also accounts for degradation due to side reactions. To address the uncertainty in the forecast, an ensemble of forecasted disturbances is generated to approximate the solution to a formulated robust optimization problem. Year-long simulation results based on real-world prosumer data show that the new method can bring a reduction of over 30\% in the average daily cost for the prosumer compared to conventional strategies based on mean nonlinear MPC. To reduce the computational complexity of the online computation required by the developed algorithm based on forecast and battery model information, one possible solution is to leverage the available data and physical information to learn the optimal dispatch schedule. Such an approach has the potential to yield fruitful results in future research.

\appendix

\begin{table}[!t]
\centering\footnotesize
\caption{Li-ion Battery Parameters}
\begin{tabular}{l c c c  c}
    \hline
     Sym.  & Unit  & $+$ & sep  & $-$\\
     \hline
     $R_p$  & m & $2\times 10^{-6}$& -- & $2\times 10^{-6}$ \\
     $D_s$  & $\text{m}^2/\text{s}$ & $1\times 10^{-14}$& -- & $3.9\times 10^{-14}$ \\
     $a_s$  & 1/m & $8.85\times 10^6$& -- & $7.236\times 10^6$ \\
    $L$ &  m  & $80 \times 10^{-6}$ &  $2.5\times 10^6$ & $88 \times 10^{-6}$ \\
    $\varepsilon_s$  & - & $0.59$ &  -- & $0.4824$ \\
     $c_{s,\text{max}}$  & mol$/$m$^{3}$  & $51,554$ &  -- & $30,555$  \\
     $\theta_{0\%}$  & - & $0.9337$ &  -- &  $0.02$ \\
     $\theta_{100\%}$  & - & $0.4855$ &  -- & $0.8608$ \\
     $k_{\text{eff}}$  & - & $2.33 \times 10^{-11}$ &  -- & $5.03 \times 10^{-11}$\\
     $F$  & C$/$mol & \multicolumn{3}{c}{$96,487$} \\
     $T_{\text{ref}}$  & K & \multicolumn{3}{c}{$298.15$} \\
     $R_{g}$  & J/(K$\cdot$mol) & \multicolumn{3}{c}{$8.314$} \\
     $c_e^0$  & mol$/$m$^{3}$  &  \multicolumn{3}{c}{$1000$}\\
     $i_{0,\text{sr}}$ & $\text{A}\cdot\text{m}^{2}$  & \multicolumn{3}{c}{Fitted $^\text{a}$} \\
     $U_{\text{sr},\text{ref}}$ & V  & \multicolumn{3}{c}{$0.21$ \cite{fu2015jps}}  \\
     $r_{f0}$ & $\Omega\cdot \text{m}^2$  & \multicolumn{3}{c}{$0.01$} \\
     $M_{f}$ & kg/mol  & \multicolumn{3}{c}{$0.162$ \cite{safari2009jes}} \\
     $\rho_{f}$ & $\text{kg/m}^3$  & \multicolumn{3}{c}{$1690$ \cite{safari2009jes}} \\
     $\kappa_{f}$ & $\text{S}/\text{m}$  & \multicolumn{3}{c}{$5 \times 10^{-6}$ \cite{safari2009jes}} \\
     $Q_{0}$ & Ah  & \multicolumn{3}{c}{$1.747$} \\
     $A$ & m$^{2}$  & \multicolumn{3}{c}{$0.0598$ $^\text{b}$} \\
   \hline
\multicolumn{5}{l}{a: linearly fitted based on $0.39 \times 10^{-7} (T_{\text{bat}} = 273.15~\text{K})$, }\\
\multicolumn{5}{l}{$2.28 \times 10^{-7} (T_{\text{bat}} = 298.15~\text{K})$, $6.3 \times 10^{-7} (T_{\text{bat}} = 323.15~\text{K})$ \cite{fu2015jps}.}\\
\multicolumn{5}{l}{b: calculated based on $Q_0 = 1.747$ Ah.}\\
     \hline
\end{tabular} \label{tab:tab_A}
\end{table}

By extending an isothermal model of Li-ion cell \cite{liyang2020apen},  $V_1^\pm$,  $R_\Sigma^\pm$,  $I_{\text{sr}}$, and $R_e$ in \eqref{eq:eq_2e}--\eqref{eq:eq_2h} can be calculated as follows.

\emph{1) Open-Circuit Potentials}
\begin{align} \label{eq:eq_a1}
{V}_1^{\pm} = {V}_{1,\text{ref}}^{\pm}({\theta }^{\pm}) + K_T^{\pm}({\theta }^{\pm})(T_{\text{bat}} - T_{\text{ref}})
\end{align}
where the reference voltages ${V}^{\pm}_{1,\text{ref}}$ and entropic coefficients $K^{\pm}_T$ are nonlinear functions of normalized Li-ion concentration ${{\theta }^{\pm}}$, and the relationship between ${{\theta }^{\pm}}$ and ${Q_1^{\pm}}$ is
\begin{align}
{{\theta }^{+}}&=\theta_{0\%}^{+}-\frac{3600}{A{{L}^{+}}F\varepsilon _{s}^{+}c_{s,\text{max}}^{+}}Q_{1}^{+}, \\
{{\theta }^{-}}&=\theta_{100\%}^{-}-\frac{3600}{A{{L}^{-}}F\varepsilon _{s}^{-}c_{s,\text{max}}^{-}}Q_{1}^{-}.
\end{align}

\emph{2) Charge-Transfer and Solid-Phase Diffusion Resistances}
\begin{equation}
R_\Sigma^{\pm }=\frac{{{R}_{g}}T_{\text{bat}}}{F}\frac{1}{A{{L}^{\pm }}{{a}_s^{\pm }}}\frac{1}{i_0^{\pm}} +
\frac{-{\partial V_1^{{\pm}}}/{\partial \theta^{{\pm}}}}{A L^{\pm} F \varepsilon_s^{\pm} c^{\pm}_{s,\text{max}}} \frac{(R_p^{\pm})^2}{15  D_s^{\pm}}
\end{equation}
where $i_0^{\pm}$  is the exchange current density, given by
\begin{align}
i_0^{\pm} = F k_{\text{eff}}^{\pm }c_{s,\text{max}}^{\pm }\sqrt{c_{e}^{0}{{\theta }^{\pm }}(1-{{\theta }^{\pm }})}.
\end{align}

\emph{3) Electrolyte Resistance}

\begin{equation}
R_e = \frac{ L^+ + 2 L^{\text{sep}}  + L^- }{2 A \kappa_{\text{eff}}},
\end{equation}
where the effective electrolyte conductivity $\kappa_{\text{eff}}$ is a function of temperature.

\emph{4) Side-Reaction Current}

The side reaction current is assumed to be kinetically controlled and governed by
\begin{equation} \label{eq:eq_Isr}
{{I}_{\text{sr}}}=A{{L}^{-}} J_{\text{sr}} = A{{L}^{-}}\frac{\alpha \beta +\alpha \sqrt{{{\beta }^{2}}+(1-2\gamma \alpha )}}{1-2\gamma \alpha }\end{equation}
where $J_{\text{sr}}$ is the side reaction volumetric current density and
\begin{align} \label{eq:eq_a4}
&\alpha =-{{i}_{0,\text{sr}}}{{a}_s^{-}}\exp \left( \frac{F(U_{\text{sr,ref}}-V_1^{-})}{2{{R}_{g}} T_{\text{bat}} } \right)  \\
 & \beta =\frac{{{I}_{\text{bat}}}}{2 A {{L}^{-}} {{a}_s^{-}} i_0^{-}} \\ &\gamma =\frac{1}{2{{a}_s^{-}} i_0^{-}}.
\end{align}

\emph{5) SEI Film Resistance}

The resistance of SEI film is calculated by
\begin{align}
{{R}_{f}}&=\frac{{{r}_{f}}}{{{a}_{s}^-}AL^-} \\
{{r}_{f}}&={{r}_{f0}}-\int_{0}^{t}{\frac{3600 {{J}_{\text{sr}}}(\tau )}{{{a}_{s}^-}F}\frac{{{M}_{f}}}{{{\rho }_{f}}{{\kappa }_{f}}}d\tau }.
\end{align}

Considering \eqref{eq:eq_Isr} and \eqref{eq:eq_5}, the initial resistance $R_{f0}$ and  the coefficient $K_f$ in \eqref{eq:eq_6} can be obtained as
\begin{align}
{{R}_{f0}}&=\frac{{{r}_{f0}}}{{{a}_{s}^-}AL^-} \\
{{K}_{f}}&=\frac{{{M}_{f}}}{{{\rho }_{f}}{{\kappa }_{f}}}\frac{3600}{F{{(A{L^{-}}a_{s}^{-})}^{2}}}.
\end{align}

The parameters of the battery model used in this work are given in Table I. Furthermore, parameters $k_{\text{eff}}^{\pm }$, $D_{s}^{\pm }$, and $\kappa_{\text{eff}}$ are functions of the temperature. All the parameters and the concentration- or temperature-dependent functions are obtained from \cite{torchio2016jes} unless stated otherwise.

\bibliographystyle{IEEEtran}%
\bibliography{root}%

\begin{thebibliography}{10}
\providecommand{\url}[1]{#1}
\csname url@samestyle\endcsname
\providecommand{\newblock}{\relax}
\providecommand{\bibinfo}[2]{#2}
\providecommand{\BIBentrySTDinterwordspacing}{\spaceskip=0pt\relax}
\providecommand{\BIBentryALTinterwordstretchfactor}{4}
\providecommand{\BIBentryALTinterwordspacing}{\spaceskip=\fontdimen2\font plus
\BIBentryALTinterwordstretchfactor\fontdimen3\font minus
  \fontdimen4\font\relax}
\providecommand{\BIBforeignlanguage}[2]{{%
\expandafter\ifx\csname l@#1\endcsname\relax
\typeout{** WARNING: IEEEtran.bst: No hyphenation pattern has been}%
\typeout{** loaded for the language `#1'. Using the pattern for}%
\typeout{** the default language instead.}%
\else
\language=\csname l@#1\endcsname
\fi
#2}}
\providecommand{\BIBdecl}{\relax}
\BIBdecl
\renewcommand{\BIBentryALTinterwordstretchfactor}{4}

\bibitem{irena2019a}
\BIBentryALTinterwordspacing
IRENA, ``Renewables 2021 global status report,'' Paris: REN21 Secretariat,
  2021. [Online]. Available:
  \url{https://www.ren21.net/wp-content/uploads/2019/05/GSR2021\_Full\_Report.pdf}
\BIBentrySTDinterwordspacing

\bibitem{csiro}
\BIBentryALTinterwordspacing
CSIRO, ``Australia installs record-breaking number of rooftop solar panels,''
  2021. [Online]. Available:
  \url{https://www.csiro.au/en/news/News-releases/2021/Australia-installs-record-breaking-number-of-rooftop-solar-panels}
\BIBentrySTDinterwordspacing

\bibitem{kong2021tsg}
W.~Kong, F.~Luo, Y.~Jia, Z.~Y. Dong, and J.~Liu, ``Benefits of home energy
  storage utilization: An australian case study of demand charge practices in
  residential sector,'' \emph{IEEE Trans. Smart Grid}, vol.~12, no.~4, pp.
  3086--3096, 2021.

\bibitem{ito2018tcst}
A.~Ito, A.~Kawashima, T.~Suzuki, S.~Inagaki, T.~Yamaguchi, and Z.~Zhou, ``Model
  predictive charging control of in-vehicle batteries for home energy
  management based on vehicle state prediction,'' \emph{IEEE Trans. Control
  Syst. Technol.}, vol.~26, no.~1, pp. 51--64, 2018.

\bibitem{gomes2022energies}
I.~Gomes, K.~Bot, M.~G. Ruano, and A.~Ruano, ``Recent techniques used in home
  energy management systems: A review,'' \emph{Energies}, vol.~15, no.~8, p.
  2866, 2022.

\bibitem{azuatalam2018rser}
D.~Azuatalam, K.~Paridari, Y.~Ma, M.~F{\"o}rstl, A.~C. Chapman, and
  G.~Verbi\v{c}, ``Energy management of small-scale {PV}-battery systems: A
  systematic review considering practical implementation, computational
  requirements, quality of input data and battery degradation,'' \emph{Renew.
  Sustain. Energy Rev.}, vol. 112, pp. 555--570, Sep. 2019.

\bibitem{bozchalui2012a}
M.~C. Bozchalui, S.~A. Hashmi, H.~Hassen, C.~A. Canizares, and K.~Bhattacharya,
  ``Optimal operation of residential energy hubs in smart grids,'' \emph{IEEE
  Trans. Smart Grid}, vol.~3, no.~4, pp. 1755--1766, Dec. 2012.

\bibitem{erdinc2015a}
O.~Erdinc, N.~G. Paterakis, T.~D.~P. Mendes, A.~G. Bakirtzis, and J.~P.~S.
  Catal\~{a}o, ``Smart household operation considering bi-directional {EV} and
  {ESS} utilization by real-time pricing-based {DR},'' \emph{IEEE Trans. Smart
  Grid}, vol.~6, no.~3, pp. 1281--1291, Sep. 2015.

\bibitem{wu2016a}
X.~Wu, X.~Hu, S.~Moura, X.~Yin, and V.~Pickert, ``Stochastic control of smart
  home energy management with plug-in electric vehicle battery energy storage
  and photovoltaic array,'' \emph{J. Power Sources}, vol. 333, no. Supplement
  C, pp. 203--212, Nov. 2016.

\bibitem{saberi2021tsg}
H.~Saberi, C.~Zhang, and Z.~Y. Dong, ``Data-driven distributionally robust
  hierarchical coordination for home energy management,'' \emph{IEEE Trans.
  Smart Grid}, vol.~12, no.~5, pp. 4090--4101, 2021.

\bibitem{nizami2018}
M.~S.~H. {Nizami}, M.~J. {Hossain}, K.~{Mahmud}, and J.~{Ravishankar}, ``Energy
  cost optimization and {DER} scheduling for unified energy management system
  of residential neighborhood,'' in \emph{Proc. IEEE EEEIC/I CPS Eur.},
  Palermo, Italy, 18 Oct. 2018, pp. 1--6.

\bibitem{mahmud2016}
K.~{Mahmud}, S.~{Morsalin}, Y.~R. {Kafle}, and G.~E. {Town}, ``Improved peak
  shaving in grid-connected domestic power systems combining photovoltaic
  generation, battery storage, and {V2G}-capable electric vehicle,'' in
  \emph{Proc. IEEE Int. Conf. Power Syst. Technol.}, Wollongong, NSW,
  Australia, 28 Sep.-1 Oct. 2016, pp. 1--4.

\bibitem{lokeshgupta2019}
B.~{Lokeshgupta} and S.~{Sivasubramani}, ``Cooperative game theory approach for
  multi-objective home energy management with renewable energy integration,''
  \emph{IET Smart Grid}, vol.~2, no.~1, pp. 34--41, Apr. 2019.

\bibitem{liyang2020apen}
Y.~Li, M.~Vilathgamuwa, S.~S. Choi, B.~Xiong, J.~Tang, Y.~Su, and Y.~Wang,
  ``Design of minimum cost degradation-conscious lithium-ion battery energy
  storage system to achieve renewable power dispatchability,'' \emph{Appl.
  Energy}, vol. 260, p. 114282, Feb. 2020.

\bibitem{hou2019}
X.~{Hou}, J.~{Wang}, T.~{Huang}, T.~{Wang}, and P.~{Wang}, ``Smart home energy
  management optimization method considering energy storage and electric
  vehicle,'' \emph{IEEE Access}, vol.~7, pp. 144\,010--144\,020, Oct. 2019.

\bibitem{sun2016a}
C.~Sun, F.~Sun, and S.~J. Moura, ``Nonlinear predictive energy management of
  residential buildings with photovoltaics \& batteries,'' \emph{J. Power
  Sources}, vol. 325, no. Supplement C, pp. 723--731, Sep. 2016.

\bibitem{dinh2020}
H.~T. {Dinh}, J.~{Yun}, D.~M. {Kim}, K.~{Lee}, and D.~{Kim}, ``A home energy
  management system with renewable energy and energy storage utilizing main
  grid and electricity selling,'' \emph{IEEE Access}, vol.~8, pp.
  49\,436--49\,450, Mar. 2020.

\bibitem{DERROUAZIN2017238}
A.~Derrouazin, M.~Aillerie, N.~Mekkakia-Maaza, and J.-P. Charles, ``Multi
  input-output fuzzy logic smart controller for a residential hybrid
  solar-wind-storage energy system,'' \emph{Energy Convers. Manage.}, vol. 148,
  pp. 238--250, 2017.

\bibitem{pozzi2020tcst}
A.~Pozzi, M.~Zambelli, A.~Ferrara, and D.~M. Raimondo, ``Balancing-aware
  charging strategy for series-connected lithium-ion cells: A nonlinear model
  predictive control approach,'' \emph{IEEE Trans. Control Syst. Technol.},
  vol.~28, no.~5, pp. 1862--1877, Sep. 2020.

\bibitem{pippia2020tcst}
T.~Pippia, J.~Sijs, and B.~D. Schutter, ``A single-level rule-based model
  predictive control approach for energy management of grid-connected
  microgrids,'' \emph{IEEE Trans. Control Syst. Technol.}, vol.~28, no.~6, pp.
  2364--2376, 2020.

\bibitem{reniers2021jps}
J.~M. Reniers, G.~Mulder, and D.~A. Howey, ``Unlocking extra value from grid
  batteries using advanced models,'' \emph{J. Power Sources}, vol. 487, p.
  229355, Mar. 2021.

\bibitem{liyang2018iecon}
Y.~Li, D.~M. Vilathgamuwa, S.~S. Choi, T.~W. Farrell, N.~T. Tran, and
  J.~Teague, ``Nonlinear model predictive control of photovoltaic-battery
  system for short-term dispatch,'' in \emph{Proc. Annu. Conf. IEEE Ind.
  Electron. Soc.}, Washington, DC, USA, 21-23 Oct. 2018, pp. 1884--1889.

\bibitem{CAI2019478}
J.~Cai, H.~Zhang, and X.~Jin, ``Aging-aware predictive control of {PV}-battery
  assets in buildings,'' \emph{Appl. Energy}, vol. 236, pp. 478--488, 2019.

\bibitem{mesbah2016mcs}
A.~Mesbah, ``Stochastic model predictive control: An overview and perspectives
  for future research,'' \emph{IEEE Control Syst. Mag.}, vol.~36, no.~6, pp.
  30--44, 2016.

\bibitem{garifi2018}
K.~Garifi, K.~Baker, B.~Touri, and D.~Christensen, ``Stochastic model
  predictive control for demand response in a home energy management system,''
  in \emph{Proc. IEEE Power Energy Soc. Gen. Meeting (PESGM)}, 5-10 Aug. 2018,
  pp. 1--5.

\bibitem{lucia2014jpc}
S.~Lucia, J.~A.~E. Andersson, H.~Brandt, M.~Diehl, and S.~Engell, ``Handling
  uncertainty in economic nonlinear model predictive control: A comparative
  case study,'' \emph{J. Process Control}, vol.~24, no.~8, pp. 1247--1259, Aug.
  2014.

\bibitem{moura2013tcst}
S.~J. Moura, J.~L. Stein, and H.~K. Fathy, ``Battery-health conscious power
  management in plug-in hybrid electric vehicles via electrochemical modeling
  and stochastic control,'' \emph{IEEE Trans. Control Syst. Technol.}, vol.~21,
  no.~3, pp. 679--694, May 2013.

\bibitem{boiroux2016ecc}
D.~Boiroux, M.~Hagdrup, Z.~Mahmoudi, K.~Poulsen, H.~Madsen, and J.~B.
  J{\o}rgensen, ``An ensemble nonlinear model predictive control algorithm in
  an artificial pancreas for people with type 1 diabetes,'' in \emph{Proc. Eur.
  Control Conf.}, 29 Jun.-1 Jul. 2016, pp. 2115--2120.

\bibitem{garcia2019acc}
J.~Garcia-Tirado, P.~Colmegna, J.~Corbett, B.~Ozaslan, and M.~D. Breton,
  ``Ensemble model predictive control strategies can reduce exercise
  hypoglycemia in type 1 diabetes: In silico studies,'' in \emph{Proc. Amer.
  Control Conf.}, 10-12 Jul. 2019, pp. 4752--4758.

\bibitem{liyang2019apen}
Y.~Li, M.~Vilathgamuwa, S.~S. Choi, T.~W. Farrell, N.~T. Tran, and J.~Teague,
  ``Development of a degradation-conscious physics-based lithium-ion battery
  model for use in power system planning studies,'' \emph{Appl. Energy}, vol.
  248, pp. 512--525, Aug. 2019.

\bibitem{liyang2019igbsg}
Y.~Li, Y.~Yang, J.~Tang, B.~Xiong, X.~Deng, and D.~Tang, ``Design of
  degradation-conscious optimal dispatch strategy for home energy management
  system with rooftop {PV} and lithium-ion batteries,'' in \emph{Proc. Int.
  Conf. Intell. Green Building Smart Grid}, Yichang, China, 6-9 Sep. 2019, pp.
  741--746.

\bibitem{rosewater2020}
D.~M. Rosewater, D.~A. Copp, T.~A. Nguyen, R.~H. Byrne, and S.~Santoso,
  ``Battery energy storage models for optimal control,'' \emph{IEEE Access},
  vol.~7, pp. 178\,357--178\,391, 2019.

\bibitem{YOO2019226715}
K.~Yoo and J.~Kim, ``Thermal behavior of full-scale battery pack based on
  comprehensive heat-generation model,'' \emph{J. Power Sources}, vol. 433, p.
  226715, 2019.

\bibitem{7984899}
M.~{Shafie-Khah} and P.~{Siano}, ``A stochastic home energy management system
  considering satisfaction cost and response fatigue,'' \emph{IEEE Trans. Ind.
  Inform.}, vol.~14, no.~2, pp. 629--638, Feb. 2018.

\bibitem{ahmedr2020rser}
R.~Ahmed, V.~Sreeram, Y.~Mishra, and M.~D. Arif, ``A review and evaluation of
  the state-of-the-art in {PV} solar power forecasting: {T}echniques and
  optimization,'' \emph{Renew. Sustain. Energy Rev.}, vol. 124, p. 109792, May
  2020.

\bibitem{wenl2020epsr}
L.~Wen, K.~Zhou, and S.~Yang, ``Load demand forecasting of residential
  buildings using a deep learning model,'' \emph{Electr. Power Syst. Res.},
  vol. 179, p. 106073, Feb. 2020.

\bibitem{marti2015cce}
R.~Mart\'{i}, S.~Lucia, D.~Sarabia, R.~Paulen, S.~Engell, and C.~de~Prada,
  ``Improving scenario decomposition algorithms for robust nonlinear model
  predictive control,'' \emph{Comput. Chem. Eng.}, vol.~79, pp. 30--45, Aug.
  2015.

\bibitem{dinh2022scirep}
H.~T. Dinh, D.~Kim, and D.~Kim, ``Milp-based optimal day-ahead scheduling for a
  system-centric community energy management system supporting different types
  of homes and energy trading,'' \emph{Sci. Rep.}, vol.~12, no.~1, p. 18305,
  Oct. 2022.

\bibitem{VAGROPOULOS20169}
S.~I. Vagropoulos, E.~G. Kardakos, C.~K. Simoglou, A.~G. Bakirtzis, and {J. P.
  S. Catal\~{a}o}, ``{ANN}-based scenario generation methodology for stochastic
  variables of electric power systems,'' \emph{Electr. Power Syst. Res.}, vol.
  134, pp. 9--18, 2016.

\bibitem{mardia1970}
K.~V. Mardia, ``Measures of multivariate skewness and kurtosis with
  applications,'' \emph{Biometrika}, vol.~57, no.~3, pp. 519--530, 1970.

\bibitem{torchio2016jes}
M.~Torchio, L.~Magni, R.~B. Gopaluni, R.~D. Braatz, and D.~M. Raimondo,
  ``{LIONSIMBA}: A matlab framework based on a finite volume model suitable for
  {Li}-ion battery design, simulation, and control,'' \emph{J. Electrochem.
  Soc.}, vol. 163, no.~7, pp. A1192--A1205, Jan. 2016.

\bibitem{MIGUEL2021103388}
E.~Miguel, G.~L. Plett, M.~S. Trimboli, L.~Oca, U.~Iraola, and E.~Bekaert,
  ``Review of computational parameter estimation methods for electrochemical
  models,'' \emph{J. Energy Storage}, vol.~44, p. 103388, 2021.

\bibitem{nrelsolardata}
\BIBentryALTinterwordspacing
M.~Sengupta and A.~Andreas, ``Oahu solar measurement grid (1-year
  archive):1-second solar irradiance,'' 2010. [Online]. Available:
  \url{http://dx.doi.org/10.5439/1052451}
\BIBentrySTDinterwordspacing

\bibitem{opsd2020household}
\BIBentryALTinterwordspacing
``Open power system data,'' 2020. [Online]. Available:
  \url{https://data.open-power-system-data.org/household_data/2020-04-15/}
\BIBentrySTDinterwordspacing

\bibitem{ameo}
\BIBentryALTinterwordspacing
``Australian energy market operator: Aggregated price and demand data,'' 2020.
  [Online]. Available:
  \url{https://aemo.com.au/energy-systems/electricity/national-electricity-market-nem/data-nem/aggregated-data}
\BIBentrySTDinterwordspacing

\bibitem{liyang2021tie}
Y.~Li, Z.~Wei, B.~Xiong, and D.~M. Vilathgamuwa, ``Adaptive ensemble-based
  electrochemical-thermal degradation state estimation of lithium-ion
  batteries,'' \emph{IEEE Trans. Ind. Electron.}, vol.~69, no.~7, pp.
  6984--6996, Jul. 2022.

\bibitem{fu2015jps}
R.~Fu, S.-Y. Choe, V.~Agubra, and J.~Fergus, ``Development of a physics-based
  degradation model for lithium ion polymer batteries considering side
  reactions,'' \emph{J. Power Sources}, vol. 278, pp. 506--521, Mar. 2015.

\bibitem{safari2009jes}
M.~Safari, M.~Morcrette, A.~Teyssot, and C.~Delacourt, ``Multimodal
  physics-based aging model for life prediction of {Li}-ion batteries,''
  \emph{J. Electrochem. Soc.}, vol. 156, no.~3, pp. A145--A153, Mar. 2009.

\end{thebibliography}

\end{document}